\documentclass[smallextended]{svjour3}       % onecolumn (second format)
\smartqed  % flush right qed marks, e.g. at end of proof
\usepackage{comment}
\usepackage{url}
\usepackage{algorithm}
\usepackage{algorithmic}
\usepackage{nth}
\usepackage{amssymb}
\usepackage{epsfig}
\usepackage{wrapfig}
\usepackage{enumitem}
\usepackage{subfig}
\usepackage{color}
\usepackage{multirow}
\usepackage[table]{xcolor}
\newcommand{\forceindent}{\leavevmode{\parindent=1em\indent}}
\usepackage{graphicx}

 \newenvironment{packed_itemize}{
 \begin{itemize}
   \setlength{\itemsep}{1pt}
   \setlength{\parskip}{0pt}
   \setlength{\parsep}{0pt}
 }{\end{itemize}}

\usepackage[utf8]{inputenc}
\usepackage[english]{babel}
\usepackage{amsmath}
\begin{document}

\title{Complex Data Analysis Using Multilayer Networks: Modeling,  Efficiency, and Versatility}
\title{Multilayer Networks For Big Data Analytics: Modeling,  Efficiency, and Versatility}
\title{Multilayer Networks For Data-Driven Analysis: Modeling,  Computation, and Versatility}
\title{Multilayer Networks For Data-Driven Analysis: Modeling,  Efficiency, and Versatility}
\title{Data-Driven Analysis: Making a Case for MLNs From Modeling,  Efficiency, and Versatility Perspective}
\title{Why Should We Use MLNs For Data-Driven Analysis? Modeling,  Efficiency, and Versatility}
\title{Making a Case for MLNs for Data-Driven Analysis: Modeling,  Efficiency, and Versatility}
\title{Data-Driven Aggregate Analysis of MLNs: Modeling,  Computation, and Versatility}
\title{Beyond OLAP: Data-Driven Aggregate Analysis of Multilayer Networks}
\title{Leveraging Novel Multilayer Network Analysis to Disparate Real-World Applications }
\title{\textbf{Complex Data Sets To Knowledge Discovery}: \textit{Modeling, Mapping Of Analysis Objectives, And Aggregate Computation Over Multilayer Networks}}
\title{\textbf{From Raw Data To Knowledge Discovery Using Multilayer Networks}: \textit{Modeling, Translating Analysis Objectives, And Their Efficient Computation}}
\title{\textbf{From Raw Data To Knowledge Discovery --} \textit{Modeling, Translation of Analysis Objectives, And Their Efficient Computation} -- \textbf{Using Multilayer Networks}}
\title{\textbf{From Base Data To Knowledge Discovery --} \textit{A Life Cycle Approach} -- \textbf{Using Multilayer Networks}}

%\title{Insert your title here %\thanks{Grants or other notes
%about the article that should go on the front page should be
%placed here. General acknowledgments should be placed at the end of the article.} }
%\subtitle{Do you have a subtitle?\\ If so, write it here}

\titlerunning{From Base Data To Knowledge Discovery Using Multilayer Networks}        % if too long for running head

%\titlerunning{Short form of title}        % if too long for running head

\begin{comment}

\author{\IEEEauthorblockN{Abhishek Santra\IEEEauthorrefmark{1},
Kanthi Komar\IEEEauthorrefmark{2}, Sanjukta Bhowmick\IEEEauthorrefmark{3} and
Sharma Chakravarthy\IEEEauthorrefmark{4}}
\IEEEauthorblockA{\IEEEauthorrefmark{1}\IEEEauthorrefmark{2}\IEEEauthorrefmark{4}IT Lab and CSE Department, University of Texas at Arlington, Arlington, Texas \\
\IEEEauthorrefmark{3}CSE Department, University of North Texas, Denton, Texas \\
Email: \IEEEauthorrefmark{1}abhishek.santra@mavs.uta.edu,
\IEEEauthorrefmark{2}kanthisannappa.komar@mavs.uta.edu,\\
\IEEEauthorrefmark{3}sanjukta.bhowmick@unt.edu,
\IEEEauthorrefmark{4}sharmac@cse.uta.edu}}

\end{comment}

\author{Abhishek Santra         
    \and
        Kanthi Komar
    \and
        Sanjukta Bhowmick
    \and
        Sharma Chakravarthy
}

%\authorrunning{Short form of author list} % if too long for running head

\institute{Abhishek Santra, Kanthi Komar, Sharma Chakravarthy \at
              IT Lab and CSE Department, University of Texas at Arlington, Arlington, Texas \\
             \email{\{abhishek.santra, kanthisannappa.komar\}@mavs.uta.edu, sharmac@cse.uta.edu}           %  \\
%             \emph{Present address:} of F. Author  %  if needed
           \and
           Sanjukta Bhowmick \at
              CSE Department, University of North Texas, Denton, Texas \\
              \email{sanjukta.bhowmick@unt.edu}
}

\date{Received: date / Accepted: date}
% The correct dates will be entered by the editor

\maketitle

\begin{abstract}
    
%\begin{abstract}
\setlength{\parindent}{5ex}

Any large complex data analysis to infer or discover meaningful information/knowledge involves the following steps (in addition to data collection, cleaning, preparing the data for analysis such as attribute elimination etc.): i) Modeling the data -- an approach for modeling as well as deriving a representation of data for analysis using that approach, ii) translating analysis objectives into computations on the model generated; this can be as simple as a single computation (e.g., community detection) or may involve a sequence of operations (e.g., pair-wise community detection over multilayer networks) using expressions based on the model, iii) computation of the expressions generated -- efficiency and scalability come into picture here, and iv) drill-down of results to interpret them or understand them clearly. Beyond this, it is also meaningful to visualize results for easier understanding. Covid-19 visualization dashboard presented in this paper is an examples of this.

This paper covers all of the above steps of data analysis life cycle using a data representation that is gaining importance with multi-entity, multi-feature data sets. First, we establish the advantages of Multilayer Networks\footnote{A Multilayer Network is a set of networks (each network termed a layer) where nodes within a layer are connected by intra-layer edges and nodes between two layers can be optionally connected using inter-layer edges.} (or MLNs) as a data model as compared to extant data models. Then we provide an entity-relationship based approach to convert the data set into MLNs, for a more precise representation of the data set. Then we outline how expected analysis objectives can be algorithmically translated to aggregate analysis expressions. Finally, we demonstrate through a set of example data sets and objectives how the expressions corresponding to objectives are evaluated using an efficient decoupling-based approach and results drilled down to obtain actionable knowledge from the data set.

Using the widely used Enhanced Entity Relationship (EER) approach for data representation, we demonstrate how to generate EER diagrams for data sets and further generate, algorithmically, MLNs as well as Relational schema for analysis and drill down, respectively. Using communities and centrality for aggregate analysis, we demonstrate the flexibility of the chosen model to support diverse set of objectives. We also show that compared to current analysis approaches, a ``divide-and-conquer" approach of MLNs is more appropriate as well as efficient, and more importantly preserves structure and semantics of the results. For this computation, we need to derive expressions for each analysis objective using the MLN model. We provide guidelines to translate English queries into analysis expressions based on keywords.

Finally, we use several data sets to establish the effectiveness of modeling using MLNs and analyze them using the proposed \textit{decoupling approach}. For coverage, we use different types of MLNs for modeling, and community and centrality computations for analysis. The data sets used are from US commercial airlines, IMDb (a large international movie data set), the familiar DBLP (or bibliography database), and the Covid-19 data set. Our experimental analyses using the identified steps validate modeling, breadth of objectives that can be computed, and overall versatility of the life cycle approach. Correctness of results are verified, where possible, using independently available ground truth. Furthermore, we demonstrate drill-down that is afforded by this approach (due to structure and semantics preservation) for a better understanding and visualization of results.


\end{abstract}

\section{Introduction}
\label{sec:introduction}

Big data analytics is predicated upon our ability to model and analyze disparate and complex data sets. Relational database management systems (or RDBMSs) have served well for modeling and \textit{querying} data sets  that needs to be managed over long periods of time. Data warehouses and On-Line Analytical Processing (or OLAP) came about to improve the querying aspects of RDBMSs using more powerful multi-dimensional analysis queries that were not possible with SQL. This evolution has continued with NoSQL systems providing alternate data models and analysis for data that were difficult (or inefficient) using RDBMSs. 

On the mining and knowledge discovery side, long-term data management is not an issue, but as the complexity of data increases, data models are needed for modeling the data in the best possible way to develop specific algorithms for mining.
Although graph models have been used for this, we see a need for more powerful data models that can capture data semantics better. We see the applicability of Multilayer Networks (or MLNs) and its analysis potential as another important step in the evolution of aggregate analysis of complex data sets.

%%SC: added to avoid Kamesh's concerns
%\sharma{4/28/21}{can we add a couple of ETL product references}

\forceindent \textit{Our focus is on the complete life cycle and automating all the steps\footnote{As there are many commercial ETL (Extract, Transform, and Load) tools available \cite{etl-1,etl-2,etl-3,etl-4,etl-5}, we do not discuss pre-processing in this paper.}, as much as possible, for knowledge discovery. Hence, we are not considering efficiency issues, benchmarks, or other approaches, such as neural networks. This helps us focus on the complete life cycle rather than comparisons of individual steps with alternatives.} 
We are also delimiting our approach, in this paper, to data sets with diverse types of entities that are defined by multiple features and interact through different relationships. Although graphs are used as the basis, the analysis and computations performed in knowledge discovery are quite different from the ones addressed in NoSQL systems, such as Neo4J~\cite{product/neo4j}. As an example, consider a data set about a group of actors and directors (as entities). Each person has some attributes associated with them, such as who they co-act with, which genre they direct, etc. (termed features.) Actors and directors can also be connected (termed relationships) if an actor is directed by a director. Implicit relationships can also be inferred between two entities if they share similar features or transitivity holds. For this data set, finding a strong group of actors and directors, where actors co-act and directors direct similar genres cannot be expressed as a query. It is a computation that requires finding communities that satisfy certain properties.

\subsection{\bf Multi-Entity and Multi-Feature Data Analysis Challenges} 

A critical question is how to model multi-entity and -feature data sets that also involve relationships among entities explicitly as well as more precisely and express \& analyze objectives clearly. Data set characteristics growing from single entity type and feature and/or relationship to multiple features and relationships leads to the following new challenges;

\begin{packed_itemize}
    \item {\underline{\em Model Expressiveness.}} With the presence of multiple types of entities and different relationships among the same- and different-types of entities, it is important to use a model that is expressive, i.e., \textit{preserves clarity of semantics as well as structure of the data\footnote{By structure and semantics preservation, we imply that additional book keeping information and mapping is not needed before and after analysis. This is needed for drill down, but is different from expansion of other attributes during drill-down for understanding the results better.}} being modeled.
    
    \item {\underline{\em Flexible Analysis Alternatives.}} Analysis objectives on multi-feature data may be specified on a subset of entities/features. For example, given a data set about actors, movies in which they have acted, and directors who have directed the movies (see IMDb data set used in this paper), the analysis can involve actors and their movie rating, or other actors with whom they work, or any combination thereof. Also, in the IMDb data set, although both actors and directors are people, they are considered separate types of entities, as they perform different roles. The {\em challenge} is to allow for flexibility of mixing and matching of entities/features for analysis, while avoiding loss of information or redundancy of computations.

    \item {\underline{\em Analysis Objectives to Expressions and Their Efficient Computation:}} Beyond modeling and flexibility of analysis, it is important to have a methodology for translating or mapping analysis objectives into computations on the model and further compute them efficiently. In this paper, we present an algorithmic approach to translate objectives to analysis expressions using computations available on the data model. For efficiency of computation, we use an approach that has been proposed for MLNs~\cite{ICCS/SantraBC17,ICDMW/SantraBC17}. Here data structures, use of parallel/distributed approaches, and the ability to scale become important.
%%%\item{\em Integrating Analysis of Different Types of Entities.} In addition to variations of the features, the data sets can also contain entities of multiple types. For example, in the IMDb dataset, although both actors and directors are people, they are considered separate types of entities, as they perform different kinds of  roles. The {\em challenge} is to combine the analysis of these two sets, such as, coupling  the clusters of actors with some characteristics (co-acting) to clusters of directors with some characteristics (direct-similar-genre).
    
    \item{\underline{\em Drill-down and Visualization of Results.}} Finally, understanding the results (knowledge discovered) is extremely important. The model and computation using the model should not mask the structure and semantics of the results. This is where we also think the MLN approach has an edge as compared to earlier approaches to this problem. Visualization is likely to become easier once structure and semantics of results are preserved. We have developed an architecture for visualizing both base data and analyzed results~\cite{ICDE2021Demo/Cowiz}. We show the results from that dashboard in this paper.
\end{packed_itemize}

\subsection{Problem Statement} 

\textit{For a given data set $\mathcal{D}$ with $\mathcal{T}$ entity types, $\mathcal{F}$ features,  and a set of analysis objectives $\mathcal{O}$, \textbf{propose} a life cycle framework that (i) generates an expressive model for the data that preserves structure and semantics, (ii) allows flexibility of selecting a combination of entities and features for analysis, (iii) algorithmically generates expressions for $\mathcal{O}$ and computes them efficiently, and (iv) supports drill-down and visualization of knowledge discovered for easier understanding.}

\forceindent Recently, a generalized framework (a model and computation using that model)  has been proposed that addresses some of these challenges. Earlier work in multi-feature data analysis (see Section~\ref{sec:related-work}) were focused either on a specific application or a specific analysis technique. In this paper, we focus on the knowledge discovery life cycle using the new approach over the MLN data model.

{\bf Overview of the Paper.} Our {\em main contribution} in this paper is to develop a complete  data analysis (or knowledge discovery) life cycle using  {\em a generalized framework for modeling \& computation on multi-feature, multi-entity data sets and show its versatility and effectiveness}. Based on the survey and comparison of the currently used techniques for modeling and analyzing multi-feature data,  we have chosen the MLN approach for modeling the data and its analysis using the {\em decoupling approach}. We have chosen the EER approach for modeling data and generating the MLN model. We also demonstrate that by combining these techniques the chosen framework can handle diverse data sets with multiple features and entity types as well as varied analysis objectives i.e., the entire life cycle. 

\subsection{Contributions and Roadmap.}
%%Given below is a overview of the approaches to modeling, analysis, mapping, and computation, that form our framework.

%%\setlength{\parindent}{2ex}
\begin{packed_itemize}
    \item {\underline{\em Expressive Modeling:}} In Section~\ref{sec:alternatives} we compare the advantages and disadvantages of several approaches for modeling of multi-feature, multi-entity data, and show why multilayer networks (MLNs) are a better alternative with many advantages. 
    
    \item \underline{\textit{Analysis Life Cycle:}} In Section \ref{sec:analysis-life-cycle}, we present the steps of the analysis life cycle from data set description and objectives to discovering knowledge corresponding to those objectives. Further,  we demonstrate how drill-down of results for visualization  can be accomplished using the same framework. 
    
    %%We posit that the model should be able to represent different features of each entity individually or in combination, as well as the relationships between entities. Moreover, the model should also be able to accommodate datasets with similar or dissimilar types of entities.  
    
    \item {\underline{\em Efficient Analysis:}} %%Many different analysis objectives can be addressed from the same dataset. Computing these objectives requires combinations of different subsets of features or integrating analysis of multiple types of entities.     
    In Section~\ref{sec:decoupling} we summarize the {\em decoupling approach} using which information about each feature/entity type is analyzed separately and then composed efficiently to obtain results for the objectives. %%Decoupling provides flexibility of combining layers in multiple ways and also leads to efficient computation.
    
    \item {\underline{\em Algorithmic Translation of Objectives:}} %%Modeling and analysis form the backbone of interpreting information from the datasets. However, a critical step is to translate real world objective into analysis expressions (similar to queries.)  For multilayer networks, some of these definitions are more complicated.  
    In Section~\ref{sec:expression-generation} we present an approach to translate analysis objectives into computation expressions using the generated MLN model characteristics and available computations.

    \item {\underline{\em Validation, Drill-down, and Visualization:}} In Section~\ref{sec:experiments}, we present drill-down and visualization of results. Where possible, we validate our experimental \textit{results} with \textit{independently} available ground truth. Further, several visualization approaches have been used in our dashboard and is capable of visualizing both base data as well as analysis results. The goal is to facilitate better understanding of the results. %%We also demonstrate the computational efficiency of our decoupling approach by comparing the time with the standard network aggregation approach.
\end{packed_itemize}

We present related work in Section~\ref{sec:related-work}. We present data sets used along with expected analysis objectives in Section~\ref{sec:data-sets}.  And conclusions in Section~\ref{sec:conclusions}.
%%setlength{\parindent}{0ex}
%To summarize, the primary contribution of this paper is validating the effectiveness of MLNs for modeling and analyzing complex data sets.

%\textcolor{blue}{I think the summary statement looks weak. Need more stronger statement}
%To {\em summarize}, 

\section{Related Work}
\label{sec:related-work}

We present related work corresponding to the steps of the life cycle in this section.

\subsection{Modeling}

\textbf{EER Modeling:} Since the Seventies, \textit{EER model}~\cite{chen1976entity} has served as  a methodology, for database design, for representing important semantic information about the real-world  applications. Relational database modeling has clearly benefited from this body of work and has motivated UML (Unified Modeling Language) for OO (Object-Oriented) design. A good EER diagram based on the data and queries to be supported, is critical and goes a long way for an error-free relational database schema. Numerous tools~\cite{erwin,erdplus,ibm,toad,dbeaver,dbschema,erstudio} have been developed for creating the EER diagram and algorithmically mapping it into relations for different commercial DBMSs.

However, with the emergence of data sets with relationships among entities and complex application requirements, such as shortest paths, important neighborhoods, dominant nodes (or groups of nodes), etc, \cite{tkde/JayaramKLYE15,DBLP:journals/dke/DasSBC20}, the relational data model was not the best choice for modeling as well as analyzing them~ \cite{PAKDD/ChakravarthyBB04}. This led to the evolution of NoSQL data models including the graph data model \cite{angles2008survey}. In many applications, such as Facebook (friendship relationship), Movies (collaborations relationship) and Twitter (follower-followee relationship), relationships needed to be modeled explicitly using the graph model. This gave access to a wide variety of analyses that were available for these data models. Recently, there has been some work in the area of graph modeling from EER diagrams, but is limited to simple attribute graphs only \cite{roy2017modeling,chen1976entity,pokorny2016conceptual,angles2018property}. 
%%However, most of these works either do not handle recursive relationships (\cite{roy2017modeling}), and weak entities \cite{de2014model} or are application-specific \cite{graves1995graph}. 
Only recently, an approach~\cite{ER/KomarSBC20} has been proposed for using the EER approach for generating MLNs using data sets and analysis objectives. This approach has been adapted in this paper.
\vspace{5pt}

\noindent \textbf{MLN Generation:} There are a number of tools developed for creating the EER diagram and these tools also translate the EER diagram developed and generate system-specific Relational schema for Oracle, DB2, etc. This makes the development of a model a bit easier although the mapping from requirements to an EER diagram is still subjective, relies on experience, and hence not unique. When it comes to mining, this modeling approach is not typically adopted. A decision to use a specific mining algorithm is based on the context and experience as well as desired objectives and if needed the data is transformed into a different format. For example, association rule algorithms use different representations of data than many other mining algorithms. However, when a graph is used as a data model, choice of nodes, edges and their labels (if needed), becomes important and there may be multiple alternative ways of creating them depending on the analysis objectives. Further, creating edges may need a similarity/proximity criteria which needs to be identified or specified.

Our approach for data analysis stems from: i) the need for analyzing the same data in multiple ways, ii) a number of aggregate analysis (e.g., community, centrality, substructure, to name a few) alternatives that can be used, and iii) need for generating expressions for analysis rather than a single computation as is typically done in mining. Hence, this approach and framework, although new to knowledge discovery, is effective as we illustrate it in Section~\ref{sec:expression-generation}. This problem requires further research as we consider our contribution in this paper to be a starting point.

\subsection{Graph and MLN Analysis}

\noindent MLNs can be classified into Homogeneous (HoMLNs), Heterogeneous (HeMLNs) and Hybrid (HyMLNs) depending upon the characteristics of entities/nodes in each layer and their connectivity to other layers. If the entities and their types are of the same type across all layers, it is a HoMLN where same entities across layers are assumed to be connected implicitly. If the entities and their types are different from layer to layer, explicit edges are used, as needed, to connect entities between two layers and these correspond to HeMLNs. Hybrid MLNs (or HyMLNs) are a combination of these two.

\textbf{Analysis of HoMLNs:} \textit{Community detection} algorithms have been extended to {HoMLNs} for identifying tightly knit groups of nodes based on different feature combinations (review: \cite{CommSurveyKimL15,CommFortunatoC09,xin2018community,magnani2019community}.) %,CommFortunatoC09}). %,xin2018community}). 
Algorithms based on matrix factorization \cite{dong2012clustering}, 
cluster expansion~\cite{li2008scalable}, 
Bayesian probabilistic models \cite{xu2012model}, 
regression \cite{cai2005mining}
and spectral optimization of the modularity function based on the supra-adjacency representation \cite{zhang2017modularity}
%and a significance based score that quantifies the connectivity of an observed vertex-layer set through comparison with a fixed degree random graph model% \cite{wilson2017community}
have been developed. Further, methods have been developed to determine \textit{centrality measures} to identify highly influential entities \cite{de2013centrality,sole2014centrality,zhan2015influence}. 
However, all these approaches \textit{analyze a MLN  by reducing it to a simple graph} either by aggregating all (or a subset of) layers that is likely lead to loss of semantics as the entity and feature type information is lost. Other approaches that consider the entire MLN as a whole result in increased complexities due to repeated traversals of individual as well as connected layers.  

Recently developed decoupling-based approaches combine partial analysis results from individual layers systematically \textit{in a loss-less manner} to compute communities~\cite{ICCS/SantraBC17} or centrality hubs~\cite{ICDMW/SantraBC17} for layer combinations. There is no aggregation of layers in this approach. Due to the ``divide and conquer" approach of decoupling, this method has been shown to be more efficient as it avoids re-computation of layer communities/hubs, and also provides flexibility of analysis. %This approach nicely leverages the algorithms developed for a single network. 
%They reduce the exhaustive analysis complexity from O($2^N-1$) (for all possible subsets of layers) to linear complexity for \textit{N} layers by composing combinations of them. %The goal of this paper is to do the same for HeMLNs.

%allows the flexible analysis of any complex dataset modeled as a homogeneous multilayer network in a cost-effective manner as one does need to to generate and separately analyze the combined networks (close to O($2^N$) in an exhaustive analysis, if the multiplex has N layers).

\noindent \textbf{Analysis of HeMLNs:} Majority of HeMLN work (reviews in \cite{shi2017survey,sun2013mining}) 
focuses on developing meta-path based methods for similarity of objects~\cite{wang2016relsim}, object classification ~\cite{wang2016text},
missing link prediction~\cite{zhang2015organizational},
ranking/co-ranking~\cite{shi2016constrained},
and recommendations~\cite{shi2015semantic}.
Few existing works propose methods to generate clusters of entities \cite{melamed2014community,sun2009rankclus}. 
Most of them concentrate mainly on inter-layer edges and not the networks themselves. Moreover, existing approaches (type-independent~\cite{LayerAggDomenicoNAL14} and projection~\cite{Berenstein2016}) do not preserve the structure or types and labels of nodes/edges without extensive  mapping and unmapping before and after computation. The type independent approach collapses all layers into a single graph keeping \textit{all} nodes and edges (including inter-layer edges) sans their types and labels. Similarly, the projection-based approach projects nodes of one layer onto another layer and uses layer neighbor and inter-layer edges to collapse two layers into a single graph with one entity type instead of two.

%In this paper, for the HoMLN community and hub detection, we apply decoupling-based approach. For HeMLN, we use a  {structure-preserving HeMLN community detection} that takes into account the combined effect of layer communities, entity types, intra- and inter-layer relationships (types). % as indicated in Section~\ref{sec:ana}.

\subsection{Drill-Down and Visualization} 
Drill-down of analysis results is critical especially for complex data which has both structure and semantics. For example, it is not sufficient to know the identities of objects in a community, but also additional details of the objects. Similarly, for a centrality hub. As we are using the MLNs as the data model, we also need to know the objects across layers and their inter-connections, if it is a HeMLN. From a computation/efficiency perspective, minimal information is used for analysis and the drill-down phase is used to expand upon, to the desired extent. Our algorithms, especially the decoupling-based, make it easier to perform drill-down without any additional mappings back and forth for recreating the structure. Our schema generation also separates information needed for drill-down (Relations) and information needed for analysis (MLNs) from the same EER diagram. It also generates needed information for the translation of objectives into expressions.

Visualization is not new either and there exists a wide variety of tools for visualizing both base data, results, and drilled-down information in multiple ways~\cite{CDC-Covid,JHU-Covid,UW-Covid}. Our focus, in this paper, is to make use of available tools in the best way possible and not propose new ones. For example, we have experimented with a wide variety of tools including, maps, individual graph and community visualization, animation of features in different ways, hovering to highlight data, and real-time data fetching and display,  based on user input as menu. Perhaps the main contribution of visualization is our architecture with clearly defined modules for analysis, visual output generation, and user-interaction. In addition to the efficiency aspects of the analysis module, We have also paid attention to efficiency of visualization creation and its access by caching pre-generated results (to avoid re-generation) and use of a hash lookup~\cite{ICDE2021Demo/Cowiz}.

%sc: added to avoid Kamesh's concerns
Finally, this paper is not about efficiency of MLN analysis as they are discussed in other places~\cite{ICCS/SantraBC17,ICDMW/SantraBC17}. Also, we are not comparing graph-based analysis with other approaches, such as neural-network based knowledge discovery to keep the paper focused on the life cycle. It is possible that expression generation and analysis could be replaced with other approaches that fit this framework.

\section{Data Sets and Terminology}
\label{sec:data-sets}

We present a subset of the data sets we have analyzed using the approach presented in this paper with their description along with analysis objectives. We have chosen data sets from different application domains to illustrate the versatility of our framework. While much larger data sets can be generated, we selected these because reliable ground truth from orthogonal sources were available for some. Due to space constraints, we are discussing a \textit{subset of analysis objectives} driven by coverage. The data sets chosen cover all types of MLNs and illustrate the generality of the framework.
\vspace{5pt}

\noindent \textbf{1. US-based Airlines: } This is a data set of six US-based airlines and their flight connections (\textit{active in February 2018}) among US cities. This information has been collected from corresponding airline sources (\cite{data/Airline/American,data/Airline/Southwest,data/Airline/Delta,data/Airline/Spirit,data/Airline/Frontier,data/Airline/Allegiant}.) 
Here all the entities are of the same type: \textit{cities}. Two cities are related if there is a \textit{direct flight} between them. The data set is characterized by single entity type (city). 
The multiple features are due to the presence of multiple airlines. A similar data set  for European carriers has been analyzed in ~\cite{cardillo2013emergence} in a different way.

\noindent\underline{Analysis Objectives.} Our analysis objectives are: i) For American, Southwest, Spirit, Delta, Frontier and Allegiant Airlines rank the top five cities, that  provide  the maximum  coverage \ref{analysis:Airline1}, %{\em classify} the airlines into major/minor carriers {\bf(A2)}; 
and  ii) predict which city (taking its population into consideration) could be selected as the next hub(s) for Allegiant Airlines to expand  its coverage and avoid competition with other airlines \ref{analysis:Airline3}.

\vspace{5pt}

\setlength{\parindent}{0ex}
\textbf{2. Bibliography Database (DBLP):} As most researchers are familiar, the DBLP data set is a publicly available information repository about computer science publications in various conferences and journals. It contains author names, their affiliation, year of publication of papers, conference/journal names, and links to the papers  \cite{data/type2/DBLP}. Clearly, there are multiple entities that can be related based on different types of relationships. 

\noindent\underline{Analysis Objectives.} Our objectives for this data set are: i) For each 3-year interval group, find the most actively publishing strong author collaboration groups \ref{analysis:DBLPHeMLN1}, ii) for each conference-based paper group, find the most popular author collaboration group and further for each of them identify their most active 3-year interval group(s) \ref{analysis:DBLPHeMLN2}, and iii) %\textit{find} the co-author groups who have published in high rank conferences \textbf{(A8)}; 
identify author collaboration groups who have published in conferences VLDB and SIGMOD, but have never published in conferences DASFAA and DaWaK~\ref{analysis:DBLPHoMLN1}.

\vspace{5pt}

\setlength{\parindent}{0ex}
\textbf{3. Internet Movie Database (IMDb):} The IMDb data set is publicly available and has information about movies, TV episodes, actor, directors, ratings and genres of the movies, etc. \cite{data/type2/IMDb}. Here the entities are of different types, such as actors, directors, movies, etc. The features can also differ since actors can be connected based on co-acting or if they have worked in movies of the same genre. %This data set will be used for demonstrating the analysis of a data set both as HoMLN and HeMLN. 
This data set can also be enriched by involving additional information about actors and directors from their social media presence, such as Facebook and Twitter. This is not elaborated due to space constraints in this paper.

\noindent\underline{Analysis Objectives.} Some of the analysis objectives for this data set are: i) Cluster actors who have acted together and have a similar average rating \ref{analysis:IMDbHoMLN1}, ii) find the groups of actors who have never acted together, but are highly rated on an average and have worked in similar genres  \ref{analysis:IMDbHoMLN2}, iii) identify genre-based groups of  actors  and  directors  having  strong collaborations \ref{analysis:IMDbHeMLN1};  %{\em refine} the groups found in A6 further to identify actors and directors who have strong collaboration and worked in highly rated movies {\bf (A7)},
 and iv) identify, for each movie rating group the genre-based most popular actor and most popular director groups. From this result, find the actor and director groups having strong collaborations \ref{analysis:IMDbHeMLN2}.

\vspace{5pt}

\noindent \textbf{4. Covid-19 Data Set:} Covid-19 data reported by New York Times\footnote{This data is collected from countries, states and health officials and verified. This includes data from The Covid Tracking Project and Johns Hopkins University. Refer to  \url{https://www.nytimes.com/interactive/2020/us/about-coronavirus-data-maps.html} for how this data is curated.} \cite{NYTimes} along with census data \cite{USCensus}, and data from other trusted sources have been used to compile a data set for the 3141 counties in the US starting from February 2020 that includes features, such as number of daily new cases, number of daily new deaths, latitude-longitude of the county seat, the mean per capita income, population density by land area, total land area, educational qualifications, traffic movement and so on.

\noindent\underline{Analysis Objectives.} The inclusion of this data set is for two purposes: i) to demonstrate how MLN analysis can be applied to data sets such as Covid and ii) to visualize  \textit{analyzed results} instead of base data as is typically done. A number of similarity analysis can be done on different desired features. In this paper, we leverage the MLN aggregate analysis approach to \textit{visualize} how Covid has \textit{spread geographically} in two arbitrarily selected periods. This can be used for understanding the Covid situation corresponding to pre and post a major event, such as Vaccination drive, Spring break, lockdowns, long holiday weekends, etc. Specifically, we want to visualize how counties with  \textbf{\textit{similar}} \textit{percentage increase in covid cases/deaths} has changed across any two user-defined disjoint periods starting from February 2020, and combine that with other demographics information. Specific objective used in the visualization is given in Section~\ref{sec:expression-generation} as \ref{analysis:Covid1} and visualization results are shown in Section \ref{sec:CovidAnalysis}.

%\end{enumerate}
\setlength{\parindent}{3ex}
Our selected data sets and analysis objectives are quite varied to illustrate the versatility of the approach. They range from analysis of finding coverage of individual airlines, clusters of co-actors/co-authors to more complicated predictions like the next planned hub of an airline, future potential teaming of actors, high quality actor-director collaborations and active periods of most popular co-authors. \textit{In addition, some of the data sets, based on the objectives, may come out either as homogeneous or heterogeneous or hybrid further depicting the capability and completeness of the modeling approach.}

\subsection{Terminology}
%%We present some important graph theory concepts  that are relevant to this paper. 
%which will help us explain how datasets are modeled into networks (or graphs) and how real world queries translate to mathematical analysis objectives. 

%Note that our datasets are both homogeneneous and heterogeneous, and the analysis questions are also very varied. We selected such diversity in datatypes and analysis objectives to
% the applicability of our modeling and analysis methods over generalized multi-featured datasets and a wide range of questions.
%In the remainder of this paper, we will refer to these datasets to illustrate for modeling and analysis techniques.

%Going forward it is necessary to discuss few  important graph theory related definitions.

\noindent {\em A Network} (or graph), $G$ is an ordered pair $(V, E)$, where $V$ is a set of vertices and $E$ is a set of edges. An edge $(v,u)$ is a 2-element subset of the set $V$. The two vertices that form an edge are said to be neighbors of each other. Here we consider graphs that are undirected (the vertices in the edge are unordered.)

{\em A Community} in a graph consists of groups of vertices that are more connected to each other than to other vertices in the graph. Several algorithms have been proposed for community detection in a simple graph. This objective is achieved by optimizing network parameters such as modularity~\cite{Clauset2004} or conductance~\cite{Leskovec08}. %The combinatorial optimization problems for community detection are NP-complete~\cite{Brandes03}.  %A large number of competitive approximation algorithms exist (see reviews in ~\cite{porter09, Fortunato2009, Xie2013}).

{\em Centrality Metrics} are used for measuring the importance of vertices. They include {degree centrality} (number of neighbors), {closeness centrality} (mean distance of the vertex from other vertices), {betweenness centrality} (fraction of shortest paths passing through the vertex), and {eigenvector centrality} (the number of important neighbors of the vertex)~\cite{Newman2010}. Choice of the metric is derived from analysis objectives. 

\section{Modeling Alternatives For Multi-Entity and Multi-Feature Data}
\label{sec:alternatives}
\noindent 
%%Multiple relationships among entities or similarity between features of different entities can be concisely expressed as networks. 
%%In recent years, network analysis has become a very popular modeling tool for analyzing large datasets of interacting entities. 
We discuss the different models using which multi-feature data can be represented as a graph and argue why using multilayer networks is a better alternative.

\subsection{Graph Alternatives}

Graphs are widely used for modeling data as rich collections of computations have been developed over the years. Their usage has become even more pronounced and important due to Internet and social networking platforms, such as Twitter, Facebook, LinkedIn. Newer systems such as Neo4J are a result of this trend. We only discuss graph alternatives in this paper due to its appropriateness for these data sets. However, we use the traditional relational model which is better suited for drill-down analysis.

\begin{enumerate}
\item \textbf{Single Graph or Monoplex:}  Here the data set is represented by a single network or graph. The vertices represent the entities and the edges represent the similarity of end points based on a feature or the dyadic relationships between them. At most one edge is assumed between nodes and labels may not be supported.

\underline{{\em Advantages.}} This way of modeling data as networks is very popular due to extensive research in this area and availability of several algorithms, such as detecting cliques, communities, centrality metrics, mining subgraphs, motifs, search, etc. %%Approximations and parallel algorithms for many of these analysis objectives also exist.
 
\underline{{\em Disadvantages.}} Single graph, however, is not best-suited for representing \textit{multiple entities and features}. Although labels can be used for different entities and features, it is difficult to combine features of different categories  (e.g., numerical and categorical), in a meaningful way as one labeled edge. The problem compounds when the entities are also of different types.  Moreover, as discussed in Section~\ref{sec:decoupling}, when  analyzing a subset of entities and/or associated feature types, separate graphs have to be generated for each such combination and analysis.

%(termed single graph or simple graph or a monoplex) models entities as nodes (including labels) and features as edges (including labels and/or weights, if present) of the graph. This model gives rise to graphs with single node types (entity types are not distinguished even when multiple node types are modeled using this approach) and single edges between nodes (either for a single feature or for a combination of features). 

%Research on graph-based modeling and analysis have been around for a long time. In addition to analysis, search and querying of graphs are also becoming mainstream~\cite{DBLP:conf/cikm/MassS14,bu2014integrating}. As the size of the data sets increases and their characteristics become more complex, efficient and scalable approaches~\cite{tkde/DasC18} are being developed to cope with the increase in data size and complexity using partitioning and parallel processing techniques. Partitioned approaches with loss-less computation and fast approximate algorithms are being investigated~\cite{rahman2013approximate}.
%\sharma{4/24/2019}{abhishek, can you add some citations above}
%%In this section, we elaborate on the modeling and analysis alternatives for data sets that can be characterized as multiple entity, feature, and relationship types.

%%In this paper, we consider \textit{three alternatives to model a complex data set using graphs}. 
%%We introduce them informally here and then define them formally in Section~\ref{sec:problem-statement}. 

%\item 
%%\textcolor{blue}{Check whether attribute or knowledge graphs are the same thing}

\item \textbf{Attribute or Knowledge Graphs:} %%The simple structure of the networks can be expanded to attribute graphs. 
Here additional features of the data sets can be represented by including node types in terms of labels (even multiple labels) and multiple edges, even self-loops, corresponding to relationships for different features. %again with or without labels and/or weights. 

\underline{{\em Advantages.}} Attribute graphs have been successfully used in
subgraph mining~\cite{tkde/DasC18,datamine/KuramochiK05,KDD/HolderCD1994}%,KDD/HolderCD1994}
, querying~\cite{tkde/JayaramKLYE15,DBLP:journals/dke/DasSBC20,dawak/DasGC16}
%,dawak/DasGC16}
 and searching~\cite{bigdataconf/HaoC0HBH15} over multi-entity types and multi-feature data sets. They capture more semantic information than simple graphs, and can handle both multiple types of features and entities.

\underline{{\em Disadvantages.}} Algorithms for some key analysis functions, such as community and centrality detection are not yet available for general attribute graphs. Hence, these graphs need to be converted to a monoplex for analysis. Although different features can be stored in the graph, for every subset of features, the analysis has to be done separately. If a subset of entities/features are used for analysis, elaborate book keeping is needed before and after the analysis to identify node/edge semantics. In other words, structure is not well-captured in this representation. %%This process can lead to redundancy of computations, particularly when the subsets have large overlaps. 

%%Other models such as Tensors have been used but is outside the scope of this paper.
%%SC: dropped for DASFAA
%%\noindent {\bf Tensors:} The adjacency matrix representation of single networks can be extended to tensors for multilayer networks. A dataset of $V$ entities and $F$ features/relationships, can be represented as a tensor, $A$, of dimensions $(V \times F) \times (V\timesF)$. The entry $A^{ia}_{jb}$ gives the connection between vertex $i$ in layer $a$ and vertex $j$ in layer $b$ of the tensor $A$.
  
 %%Another popular method for representing multi-featured data is by using tensors{\em Advantages.} Many network features such as centrality, community using modularity maximization, clustering coefficients can be defined and computed as tensor operations~\cite{Domenico15}. Tensor operations are generally easier than graph based operations to optimize and parallelize for large datasets.
 
%%{\em Disadvantages.} As with the other models, tensors also do not allow for flexible composition of different features, except by analyzing each combination separately. Moreover, tensors are generally used for modeling datasets of one single type of entities and therefore are typically not applied to datasets with multiple types of entities.
 
\item \textbf{Multilayer Networks: }
\noindent Given the pros and cons of the above options, we propose modeling multi-feature, multi-entity type data sets, as multilayer networks (MLNs). Informally, MLNs are layers of single graphs (or monoplexes).
%%%\footnote{The terminology used for variants of multilayer networks varies drastically in the literature and is not even consistent with one another. Please refer to~\cite{MultiLayerSurveyKivelaABGMP13} which provides a comprehensive comparison of terminology used in the literature, and their differences clearly.}
Each layer, typically, captures the semantics of one particular feature along with associated entities. As in a monoplex, the graph vertices represent the entities of the data set and the edges represent similarity between the feature values or the dyadic relationship between the end point vertices. The vertices of two layers can also be connected. To differentiate, we term the edges within a layer as \textit{intra-layer edges} and the edges across the layers as \textit{inter-layer edges}. 

\vspace{5pt}

There exist, primarily, two distinct types of multilayer networks -- {\em homogeneous} and {\em heterogeneous}. If each layer of a MLN has the \textbf{same set of entities of the same type or nodes}, it is termed a Homogeneous MLN (or HoMLN.) For a HoMLN, intra-layer edges are shown explicitly and inter-layer edges are not shown, as they are implicit. As an example, the US-Airlines data set can be modeled using HoMLN. The nodes in each layer are the same (cities) and edges correspond to the flights between cities. Each layer captures a  different airline. Within a layer, two nodes (cities) are connected if there is a direct flight between them for that airline. It is also possible to capture additional information, such as distance, number of flights per week using edge labels in this model. Modeling of this data set using the EER diagram and the generation of MLN for the US-Airlines is discussed in Section~\ref{sec:analysis-life-cycle}.
%%Figure~\ref{fig:MLN-example}(a) shows the HoMLN example for the Airline data set\footnote{An algorithmic approach to model MLNs using data set and analysis objectives is detailed in~\cite{ER/KomarSBC20}. We show the application of this algorithm to US-airlines data set in Section~\ref{sec:analysis-life-cycle}.}.

%\asantra{3/13}{figure commented. needs to be revised with relations}

\begin{comment}

\begin{figure}[h]
\centering
%\vspace{-20pt}
\includegraphics[width=0.7\columnwidth]{figures/MLN-example-v3}
%\vspace{-20pt}
\caption{Homogenenous and Heterogenenous MLNs}
%\vspace{-20pt}
\label{fig:MLN-example}
\end{figure}

\end{comment}

\vspace{5pt}

When the \textbf{set and types of entities are different across layers}, then the MLN is termed as a heterogeneous multilayer network (HeMLN). IMDb data set is an example which generates a HeMLN. 
%%sing EER approach on the IMDb data set and the generation of heterogeneous MLN is discussed in Section~\ref{sec:analysis-life-cycle}. 
Each layer has a different entity type as its nodes (e.g., actors, directors, and movies). The graph of a layer is defined with respect to the chosen features and entity types.
In this case of HeMLNs, the inter-layer links are defined explicitly based on feature semantics that correspond to an edge (e.g., directs-actor, directs-movie, acts-in-a-movie). %This type of MLN where each layer has a different set and type of nodes is termed a \textbf{He}terogeneous \textbf{M}ulti\textbf{L}ayer \textbf{N}etwork (or \textbf{HeMLN}.) 
It is also possible that a data set generates a combination of the above two, termed a hybrid multilayer network (or HyMLN.)

%Figure~\ref{fig:MLN-example}(b) shows an example from the IMDb data set. It is possible to model the same data set in both ways based on analysis requirements. 

\textit{Note that whether a data set generates a HoMLN or HeMLN or HyMLN depends on the data set description and objectives being analyzed. Our choice of four different data sets is to showcase the effectiveness of the approach for generating appropriate MLN type needed for the analysis.}

\vspace{5pt}
\underline{\em Advantages.} Compared to the other options, multilayer networks are a more expressive and elegant for modeling data sets with multiple entities, features, and relationships. In MLNs each \textit{chosen} feature (or combination) is modeled in a separate layer and thus this model can support both heterogeneous and homogeneous data sets.
MLNs are also better suited from an information representation (i.e., \textit{structure and semantics}) viewpoint and its visualization. Instead of cluttering all the entities and relationships in a single graph (or layer), they are logically separated and hence are easy to understand. The intra- and inter-layer relationships are also separated semantically. Each 
incremental change to each feature or relationship, as modeled by addition/deletion of vertices and edges can be easily included  without extensive re-modeling of the already created MLN.
Unlike most currently used approaches there is no need to convert a MLN representation to another one (simple or attributed) for analysis when the decoupling approach, discussed in Section~\ref{sec:decoupling}, is used. Finally, a subset of the layers can be analyzed making this model flexible from a selective analysis perspective.

%Monoplexes and attribute graphs require the information of the different features to be aggregated or collapsed before analysis can be accomplished (as explained in Section ~\ref{sec:modeling}).  

%%\sharma{8/18/2019}{this is weak. need to separate disadvantages as: inability to CURRENTLy process due to aggregation/collaps with references. then bring out the challenges which is addressed in this paper. Highlight that both structure and semantics is lost in earlier approaches. the challenge is to come up with an approach that preserves both!}
%%\textcolor{red}{I tried to address it in the para below.}
%\sharma{8/19/2019}{revised, moved around}

\vspace{5pt}
\underline{\em Challenges:} Having argued for MLNs for modeling, the primary challenge is to develop new algorithms for MLNs for performing analysis. This needs to be done preserving the MLN structure and semantics, as much as possible, during analysis (i.e., without collapsing them as is done by current approaches). The difficulty with the alternative approaches that collapse is the reconstruction of final analysis results to understand them clearly. This requires two mappings one before collapsing and one for reconstruction.  This adds additional computation which is separate from drill-down. Semantics preservation is certainly needed for drill-down of results as shown in Sec.~\ref{sec:experiments}. Decoupling-based algorithms used in this paper, by definition, preserve both structure and semantics (removing mapping of entities and features back and forth) making it easier for drill-down analysis. Preservation of structure and semantics also facilitates visualization clarity of results directly. %%We address this challenge by applying network decoupling. 
Both computing community and centrality \textit{efficiently} computation algorithms used in this paper keep the MLN structure intact.

%%%The primary challenge in analyzing MLNs is that most current analysis approaches  transform the MLN models to a monoplex  or an attribute through aggregation of the edges in the layers or projections across the layers (we provide  details in Section~\ref{sec:decoupling}). Although,  some analysis is possible using this approach, the structure and semantics of the original MLN is obscured. Thus, separation of features and node types are difficult. Furthermore, individual (or aggregate) computations are difficult as most of the extant algorithms deal only with simple graphs. There are also no algorithms for community and hub detection for attribute graphs.  In the next section we present an analysis technique that improves over the current approaches to maintain the structural information while allowing for efficient analysis.

\end{enumerate}

\section{Data Analysis Life Cycle}
\label{sec:analysis-life-cycle}

\begin{figure}[h]
\centering
%\vspace{-10pt}
\includegraphics[width=\columnwidth]{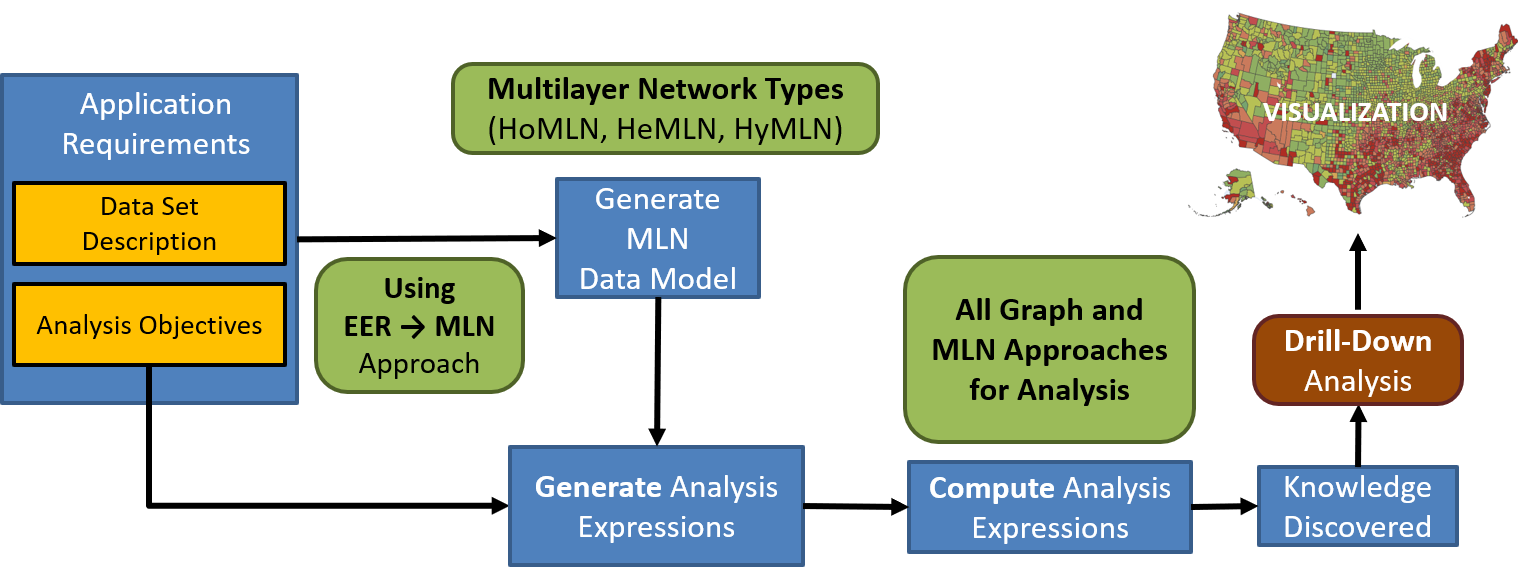}
%\vspace{-20pt}
\caption{Data Analysis or Knowledge Discovery Life Cycle}
%\vspace{-10pt}
\label{fig:life-cycle-flow}
\end{figure}

Figure~\ref{fig:life-cycle-flow} shows the steps of the data analysis life cycle from gathering base data and analysis objectives to discovery of knowledge, its drill-down, and visualization. Given a data set along with its description and the desired analysis objectives, the first step is to choose a model for representing the data. For this paper, we have chosen the multilayer network as the data model, based on the discussion in Section~\ref{sec:alternatives}. This will be generated from the data set and the analysis objectives using the approach presented in Section~\ref{sec:eer-to-mln}.

Once the model is generated, as shown in the figure, expressions for analysis objectives are generated using the MLN model characteristics (along with available computations for the chosen model) and only the analysis objectives. This is an important step and is currently done using keywords and description of the analysis objectives based on some heuristics. Details of the model, understanding of the objectives as well as the computations available on this model are important for this step. Eventually, this step needs to be automated, as much as possible, using natural language processing of the objectives along with the model characteristics and heuristics. Following this step is the actual computation of the generated expressions to discover knowledge. Ideally, any available algorithms (for graph and/or MLN in our case) can be used for this purpose. Use of model characteristics and algorithms as input, the approach used in this paper to generate the expressions are described in Section~\ref{sec:expression-generation}. The results obtained need to be drilled down further using additional attributes that were not retained as part of the MLN model. This is done using additional attributes specified in the EER diagram that are mapped to a relational model from the same EER diagram\footnote{It is not necessary to use a relational model for this as other alternatives are possible. We have used Oracle for drill-down and hence we generate the relational schema.}. For example, if a director id is used in the MLN models, details of the director in terms of location, number of movies directed etc. may be needed for drill-down. Hence, capturing all information needed for analysis and drill-down is extremely important as part of the EER diagram creation step. We will illustrate the results of detailed drill-down in Section~\ref{sec:experiments}. Even a better way is to visualize the drilled down results to make it is easier for a non-technical person to understand and interpret the analysis results. This is illustrated in Section~\ref{sec:CovidAnalysis}. 

\subsection{EER $\rightarrow$ MLN Mapping}
\label{sec:eer-to-mln}

\noindent Enhanced/Extended Entity Relationship (EER) approach is widely used for modeling data (and functionality desired) from which, typically, a database schema is derived. This has been used for all three types of traditional databases -- hierarchical, network, and relational. This modeling technique predates the UML (Unified Modeling Language) widely used today for modeling object-oriented applications as well as algorithmic flow and activity diagrams. The purpose of the EER approach is to convert data requirements gathered during the knowledge acquisition phase to develop a more precise and unambiguous model/representation that can be used for algorithmically generating the MLN schema in our case. We have used the same underlying principles for creating an EER diagram of the data set and analysis objectives (queries are used, instead, in the database context)  from which a MLN as well as a relational schema are generated. MLN schema/model is used for deriving expressions for knowledge discovery corresponding to the objectives while the relational schema/model is used for drill-down of the knowledge discovered. Once the EER diagram is generated, generation of MLN is done algorithmically. The details of motivation and an algorithm to translate an EER diagram to a MLN can be found in~\cite{ER/KomarSBC20}. Below, a brief overview is provided through examples using the data sets and analysis objectives described in Section~\ref{sec:data-sets}. For generating EER diagrams, typical heuristics used are: nouns as entities, verbs as relationships, and adjectives as attributes. Other considerations based on objectives (e.g., coverage of objectives in our case) are also taken into account.

\subsubsection{EER Modeling of the US-Airlines Data Set}

\noindent Creating an EER diagram for the US airline data set is relatively straight forward. Although the data set contains airports and their unique codes along with flight information between airports, for simplicity of analysis\footnote{There may be multiple airports associated with a city. For example, DAL and DFW are two major airports associated with the Dallas-Fort Worth metroplex.}, we have used cities instead of airports. Other information such as flight number, number of flights per week are also be available. The data set contains \textit{direct} flight information for American, Southwest, Spirit, Delta, Allegiant, and Frontier, that were active in February 2018. The analysis objectives are:

  \begin{figure}[h]
    \centering
 \vspace{-10pt}
    \includegraphics[width=0.9\linewidth]{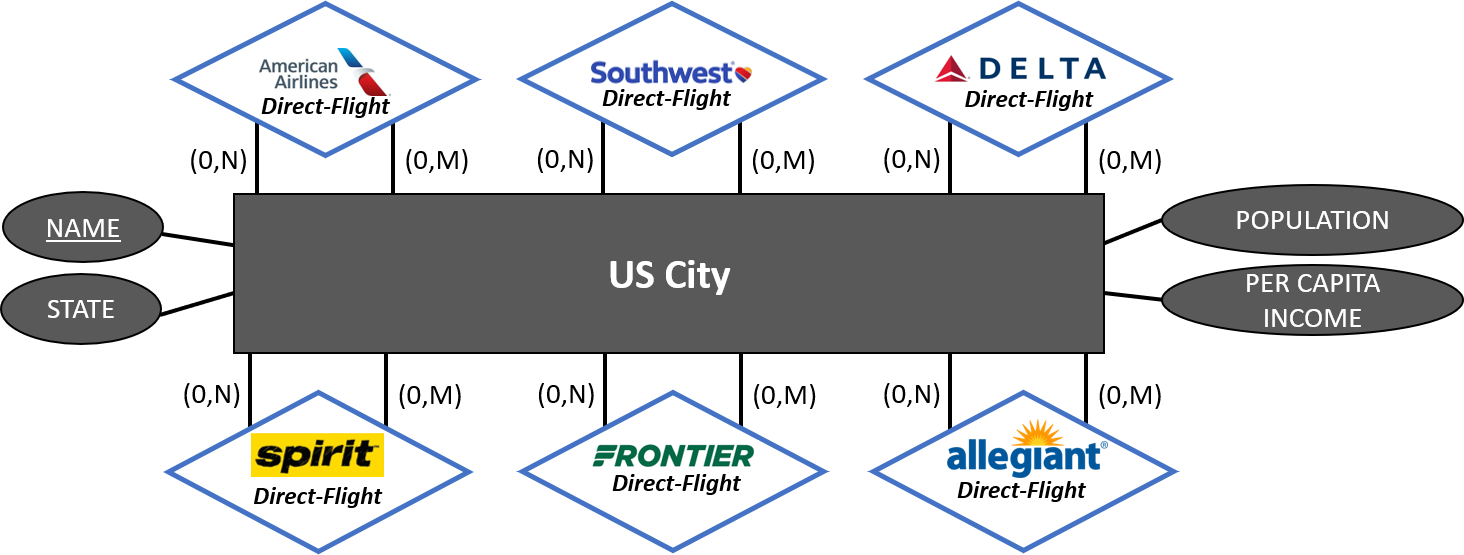}
%%\vspace{-10pt}
    \caption{US-Airlines EER Diagram}
    \label{fig:USAirline-ER}
\vspace{-10pt}
\end{figure}

  \begin{enumerate}[label={\textbf{(A\arabic*)}}]  
    \item \label{analysis:Airline1} %For each airline in the data set, \textit{Identify the top five cities for each airline}  that  provide  the maximum  coverage for that airline.
    For American, Southwest, Spirit, Delta, Frontier and Allegiant Airlines rank the top five cities, that  provide  the maximum  coverage.
%    \item \label{analysis:Airline2} Classify the airlines into major and minor carriers
    \item \label{analysis:Airline3} %Predict which city would be selected as the \textit{next hub(s)} for Allegiant Airlines to expand its coverage such that the expansion cost and competition with other airlines is minimized. \textit{Order} these predicted cities based on the population they will be catering to. 
    Predict which city (taking its population into consideration) could be selected as the next hub(s) for Allegiant Airlines to expand  its coverage and avoid competition with other airlines. 
    %%\textit{This objective can be specified for any other airline as well.}
  \end{enumerate}

  \begin{figure}[h]
    \centering
%%\vspace{-20pt}
    \includegraphics[width=0.9\linewidth]{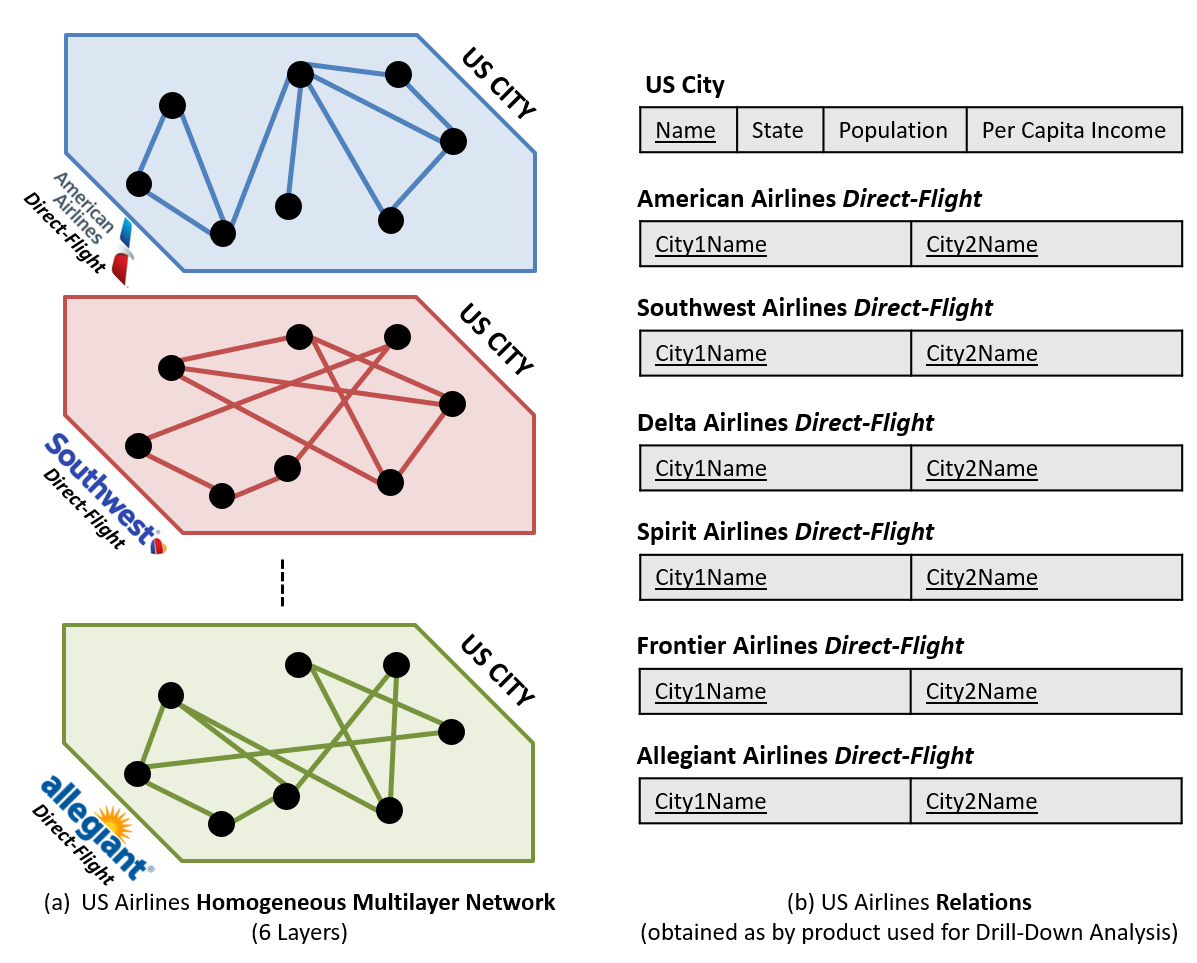}
%%\vspace{-10pt}
    \caption{MLN and Relational Schema Generated for the US-Airlines EER Diagram shown in Figure \ref{fig:USAirline-ER}}
    \label{fig:USAirline-Model}
%%\vspace{-30pt}
\end{figure}

Since  the objectives are to analyze \textit{each airline} for maximum coverage, it is clear that each airline needs to be modeled separately in the EER diagram. For that, \textit{US City} can be modeled as an entity. The direct airline flights are modeled as self-relationship between cities connecting them. A relationship is used for each airline in the EER diagram - \textit{American Direct-Flight, Southwest Direct-Flight, etc.}. The resulting EER diagram is shown in Figure~\ref{fig:USAirline-ER}. Since the analysis objectives are for these airlines, only the cities that are served by all the airlines are considered. If an individual airline is analyzed, all cities served by that airline can be included. The objectives also indicate the need for additional information for each hub for objective~\ref{analysis:Airline3} as only the hub information is not sufficient. Additional information about entities (e.g., population, per capita income etc.) is modeled as attributes of relation US City (see Figure~\ref{fig:USAirline-Model} (b)) that will be used for drill-down analysis and ranking of cities as will be shown in Section~\ref{sec:experiments}.
%%since both the nodes and edges between them are \textit{explicitly} defined. We model each layer to correspond to a specific airline. We have selected 6 airlines (layers) for analysis -- American, Southwest, Spirit, Delta, Allegiant, and Frontier. Each node in a layer represents a US city. The same set of cities are taken for each airline. Two cities are connected if there is a \textbf{direct flight} between them (active in February 2018). In Table~\ref{table:USAirlineHoMLNStats} we give the MLN statistics. This MLN is a homogeneous multilayer network. 

When the algorithm given in ~\cite{ER/KomarSBC20} is applied, the 6 layer homogeneous MLN shown in figure \ref{fig:USAirline-Model} (a) is generated. 
%The number of cities (nodes) served by all airlines under consideration and their flights (edges) are shown in Table~\ref{tab:USAirlineHoMLNStats}. 
In addition, the relations shown in figure \ref{fig:USAirline-Model} (b) are also generated for drill-down analysis and additional computations, if any.

\subsubsection{EER Modeling of the DBLP Data set}

%\sharma{3/17/21}{why are some double lines and some single? we ONLY need single lines with (min, max) cardinality info!}

\noindent Let us consider the DBLP data set along with analysis objectives described in Section~\ref{sec:data-sets}. An EER diagram for that is constructed as follows. For DBLP, the data set consists of all publications from VLDB, SIGMOD, ICDM, KDD, DaWaK and DASFAA from  2001 to 2018. The analysis objectives are as follows\footnote{Note that some of the objectives indicate specific analysis requirements, such as 3-year periods and others are stated in general terms, such as collaboration groups. These essentially are parameters associated with the objectives that need to be resolved for generating layers prior to analysis. \textit{Hence, these are modeled as parameters of the relationships whose values are needed for creating the layer graph prior to actual analysis, but are not needed for generating the MLN schema.} This is important as these parameters provide a way to perform new analysis on the same data set without changing the model and expressions  generated. Layers that are affected by the parameters need to be re-generated. This adds significant advantage for the breadth of analysis supported by the approach presented in this paper.}:

\begin{figure}[h]
    \centering
%% \vspace{-20pt}
    \includegraphics[width=\linewidth]{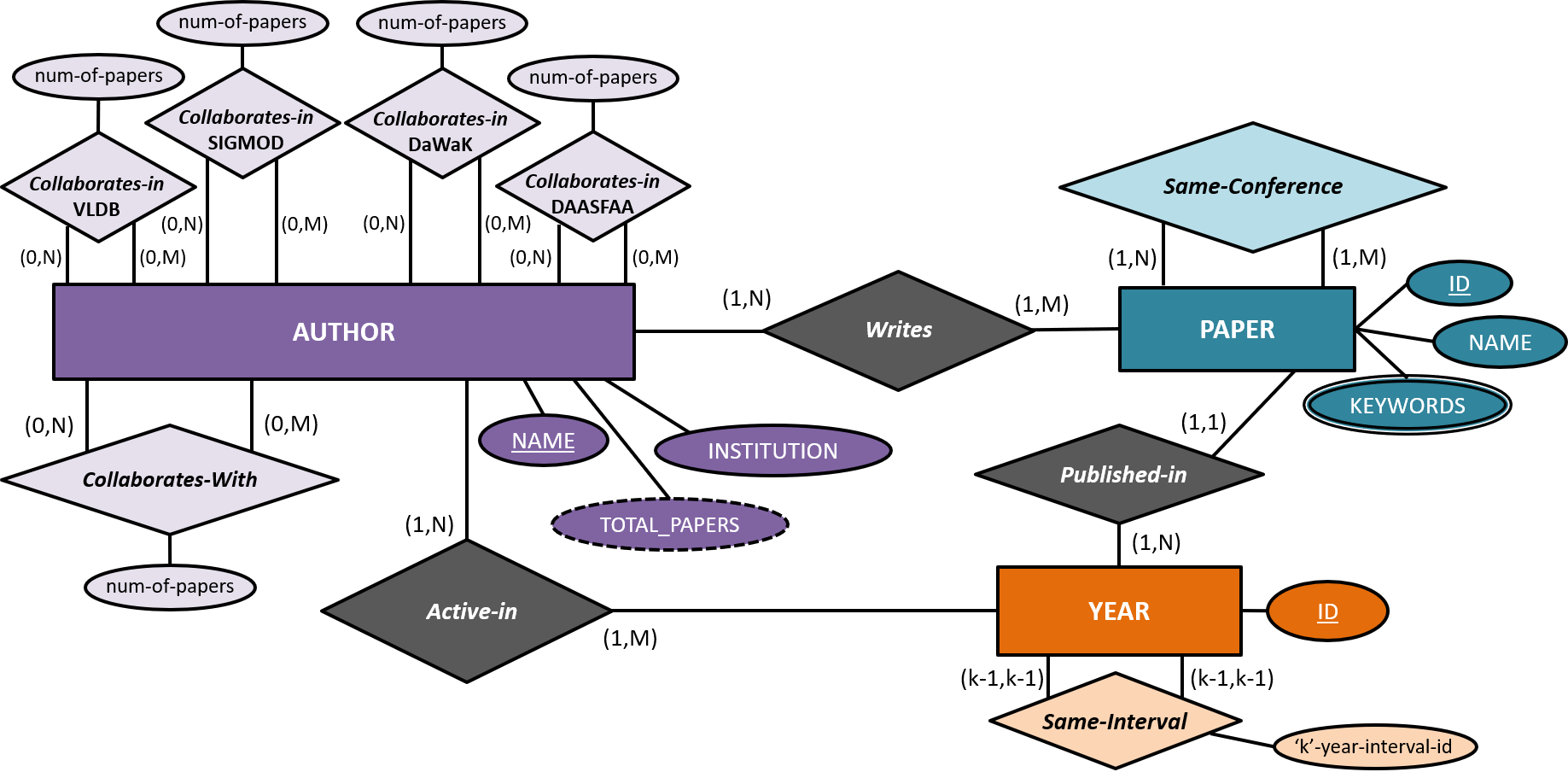}
%%\vspace{-10pt}
    \caption{DBLP EER Diagram for Objectives~\ref{analysis:DBLPHeMLN1} to~\ref{analysis:DBLPHoMLN1}}
    \label{fig:DBLPERDiagram}
\vspace{-20pt}
\end{figure}

\begin{enumerate}[label={\textbf{(A\arabic*)}}, resume]  

        \item \label{analysis:DBLPHeMLN1} %For each disjoint 3-year period, find the strong co-author groups who were most actively publishing.
        %For each disjoint 3-year period, find the strong co-author groups who were most actively publishing.
        For each 3-year interval group, find the most actively publishing strong author collaboration groups.
    \item \label{analysis:DBLPHeMLN2} %For the most popular collaborators in each conference, identify the 3-year period(s) when they were most active.
    %Conference-wise for the most popular collaborators, identify the 3-year period(s) when they were most active.
    For each conference-based paper group, find the most popular author collaboration group and further for each of them identify their most active 3-year interval group(s).
        \item \label{analysis:DBLPHoMLN1}  %Identify collaboration groups who have published in conferences VLDB and SIGMOD, but have never published in conferences DASFAA and DaWaK.
        %Identify collaboration groups who have published in conferences VLDB and SIGMOD, but have never published in conferences DASFAA and DaWaK.
        Identify author collaboration groups who have published in conferences VLDB and SIGMOD, but have never published in conferences DASFAA and DaWaK.
      \end{enumerate}  
%\sharma{3/24/2021}{check whether my revisions are ok}

Based on data set description and analysis objectives (\ref{analysis:DBLPHeMLN1} - \ref{analysis:DBLPHoMLN1}), the EER diagram shown in Figure \ref{fig:DBLPERDiagram} has been developed by following the same steps used in the previous example: i) identifying Entities, ii) identifying Relationships, including self and binary relationships, and iii) Cardinality information. \textit{Author}, \textit{Paper} and \textit{Year} come out as three entities, where some entity characteristics, such as   \textit{institution}  and \textit{keywords} are modeled as attributes of \textit{Author} and \textit{Paper} entities, respectively. \textit{Total\_Papers} attribute of (\textit{Author}) entity is shown as a derived attribute as it can be calculated using \textit{writes} binary relationship. {\em Collaborates-with}, {\em Same-Conference} and {\em Same-Interval} are the three self relationships that relate two \textit{authors} if they have worked together on papers,
%on at least 3 papers published in \textit{any} conference
two \textit{papers} if they are published in same conference, and two instances of \textit{years} if they are in the same disjoint interval, respectively.

\begin{figure}[h]
    \centering
%%   \vspace{-26pt}
    \includegraphics[width=\columnwidth]{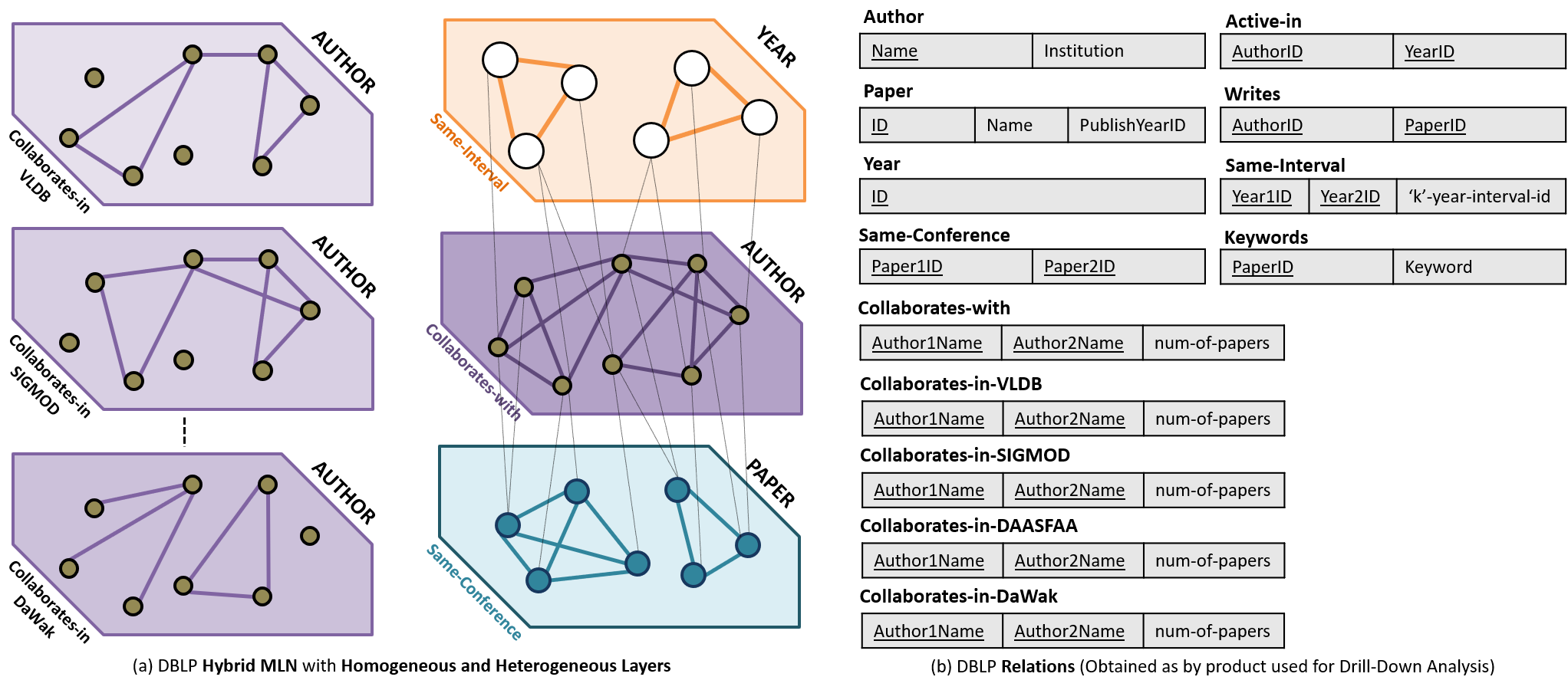}
%%    \vspace{-20pt}
    \caption{Result of EER $\rightarrow$ MLN algorithm on the DBLP EER Diagram shown in Figure \ref{fig:DBLPERDiagram}}
    \label{fig:DBLP-MLN}
%%       \vspace{-25pt}
%%\end{wrapfigure}
\end{figure}

The \textit{Collaborates-with} and \textit{Same-Interval} self relationships are associated with the attributes  \texttt{num-of-papers} and \texttt{`k'-year-interval-id}, respectively, to capture the parameters implicitly specified in the objectives. The value of these relationship attribute parameters become the basis for relating two entities and connecting two nodes in the MLN layer graph.  %Based on the analysis requirement of 3-year periods, the value of  'k' in 'k'-year-interval-id becomes 3. Moreover, two authors are related

In addition to the \textit{Collaborates-with} relationship, the author entity is associated to 4 other self relationships - \textit{Collaborates-in-VLDB}, \textit{Collaborates-in-SIGMOD}, \textit{Collaborates-in-DASFAA} and \textit{Collaborates-in-DaWak}, that capture collaboration relationship between two authors for specific conferences as required by objective \ref{analysis:DBLPHoMLN1}. Even these have the attribute parameter, \texttt{num-of-papers}, that defines the implicit notion of collaboration in the objective.  Since \textit{Author} and \textit{Paper} are distinct entities by definition, a binary relationships is needed to capture paper authorship which is a many-to-many relationship. Hence, {\em Writes}, and similarly {\em Active-in} and {\em Published-in} binary relationships capture the information to indicate if an author has written a paper, whether an author was actively publishing in a year and in which year a paper was published, respectively. Finally, the data characteristics and intuitive assumptions have been used to deduce the min-max cardinality.

\begin{comment}
    \item \textbf{(Min, Max) Cardinality Ratios:} 
    \begin{itemize}
        \item {\em Collaborates-with} recursive relationship has cardinality ratio as \textit{ (0,N)..(0,N)} as each author can work individually or with any number of authors. {\em Same-Conference} has cardinality \textit{(1,N)..(1,N)} as many papers are published in the same conference, thus a paper is related to at least one paper. Cardinality of {\em Same-Range} is \textit{(2,2)..(2,2)} as each year is related to the other 2 years in the 3-year period. \textit{Same-Score} has \textit{(0,N)..(0,N)} cardinality as a review may not be related to any other review.
        \item Binary relationship {\em Writes} between author and paper entity has \textit{(1,N)..(1,N)} cardinality as an author can publish one or more papers and also paper can have one or more authors. Similarly, {\em Active-in} has \textit{(1,N)..(1,N)} cardinality as an author is active in at least one year and in a given year many authors can be active. The {\em Published-in} relationship has \textit{(1,1)..(1,N)} cardinality as paper is published only in one year but many papers can be published in a year. Finally, for \textit{Receives} the cardinality is \textit{(3,5)..(1,1)} as every paper receives 3 to 5 reviews, however each review is for exactly one paper.

\end{itemize}
\end{comment}

%%\begin{wrapfigure}{l}{0.6\columnwidth}

Once the EER diagram is developed using the data set and analysis objectives, the algorithm in ~\cite{ER/KomarSBC20} is used to generate the MLN schema shown in Figure~\ref{fig:DBLP-MLN} (a) which happens to be Hybrid. The expression generation phase will demonstrate how this hybrid MLN will be used appropriately for expression generation of the objectives. The five self-relationships with the Author entity (\textit{Collaborates-With, Collaborates-in-VLDB, Collaborates-in-SIGMOD, Collaborates-in-DASFAA and Collaborates-in-DaWaK}) are mapped to five \textit{homogeneous} AUTHOR layers. For generating these layers, two authors are connected if they have collaborated on at least 3 papers, that is  \texttt{num-of-papers} parameter value is at least 3. The other two self-relationships (\textit{Same-Interval, and Same-Conference}) are mapped to two layers - YEAR-Same-Interval and PAPER-Same-Conference. Based on the requirement of analyzing 3-year periods, two year nodes are connected in the layer graph if they have same value of \texttt{`k'-year-interval-id} attribute parameter, where k = 3. For instance, for the years 2001 to 2018 present in the data set, [2001-2003] is interval 1 and [2016-2018] is interval 6. The binary relationships \textit{Writes, Active-in and Published-in} correspond to inter-layer edges between the corresponding layers representing the entities. %The \textit{writes} binary relationship give rise to inter-layer edges between the author layer and the other layers.
Few inter-layer edges are not illustrated in the figure to maintain the visual clarity of the figure. 
Additionally, using the same EER diagram, a Relational schema is also generated as shown in Figure \ref{fig:DBLP-MLN} (b). These are used for drill-down analysis.

For the rest of the data sets used in this paper, we will only show the MLN schema generated for them in Section~\ref{sec:expression-generation} and not go into the details of the EER modeling due to space constraints. A similar process as illustrated above is used.

\section{Analysis Alternatives For Multilayer Networks (MLNs)}
\label{sec:decoupling}

Since there are not many algorithms available for MLNs and there are a number of widely-used algorithms for single graphs for community, centrality, and substructure discovery, current approaches to MLN analysis take advantage of this by converting a MLN into a single graph.

%%We give a brief overview of the current techniques to analyze MLN and discuss how our proposed network decoupling approach improves over the traditional methods.
%%Once the data set has been modeled using one of the MLN alternatives, the goal is to use efficient algorithms for computing communities and hubs on individual layers and compose them as needed for meeting analysis objectives.

%%\footnote{In this paper, we delimit ourselves to \textit{communities} in HeMLNs. Other types of analysis, such as centrality, subgraph mining, even querying and search are applicable.}.

\subsection{Single Graph Approaches} 

The basic idea is to map the multilayer networks to an equivalent single graph in various ways~\cite{Boccaletti20141,MultiLayerSurveyKivelaABGMP13}. However, through this process, many of the information in the multilayer graphs can be lost, \textit{if appropriate mappings are not created and used. In some cases mappings can become fairly complicated}. There are mainly two approaches for converting a MLN into a single layer network. The first, used for homogeneous MLNs, is to {\em aggregate the edges of the  multilayer network}. Specifically, given two vertices $v$ and $u$, the edges between them from each layer are aggregated to form a single aggregated edge. This process is repeated for all the vertex pairs. Some typical aggregation functions are Boolean AND (intersection), OR (union) or linear functions when the edges are weighted. An example, from homogeneous MLNs, would be aggregating routes of different airlines~\cite{cardillo2013emergence} by applying OR (union). This will give rise to multiple edges between nodes or a single edge (if desired from analysis perspective). Mapping has to capture this information clearly and used before and after analysis.

For heterogeneous MLNs, aggregation is performed in many ways. The first is {\em type independent}~\cite{LayerAggDomenicoNAL14}, that is ignore the different types of the entities (and labels) present, and essentially treat it as a homogeneous MLN with a subset of vertices in each layer. The second method is  {\em projection-based} ~\cite{Berenstein2016}. Here, if two vertices in a layer are connected to a common vertex in another layer, then an edge is inferred between them.
Such ``projections" of one layer onto another layer produce inferred edges and then these edges are aggregated. %%An example is connecting drugs that act on common proteins~\cite{Berenstein2016}. 

A third approach, used for HeMLNs, is to {\em transform the multilayer network into an attribute graph}, where the vertices and edges are labeled based on their types. This graph is analyzed to find specified subgraphs, such as patterns of authors, papers and venues~\cite{sun2013mining} or vulnerabilities in infrastructure networks~\cite{Banerjee2016}. 

\underline {\em Issues.}
Single graph/network approach has the advantage that many analysis algorithms for community and hub detection are available (e.g., Infomap~\cite{InfoMap2014}, Louvain~\cite{DBLP:Louvain} being prominent ones for community detection). However, the aggregation approaches preserve neither structure nor semantics of MLNs (without explicit mapping and unmapping) as they aggregate layers. Importantly, aggregation approaches are likely to result in some information loss or distortion of properties~\cite{MultiLayerSurveyKivelaABGMP13} or hide the effect of different entity types and/or different intra- or inter-layer relationship combinations~\cite{DeDomenico201318469}. 
In cases, where the multilayer network is converted to an attribute graph, algorithms for aggregate computations (e.g., community, hub) do not exist. Again, they have to be separated into simple graphs (with at most one edge between any pair of nodes) for analyzing. This adds not only additional cost, but the purpose of modeling is defeated to  a large extent. Some approaches use the multilayer network as a whole~\cite{Wilson:2017:CEM:3122009.3208030,magnani2019community} and use inter-layer edges, but do not preserve the layer semantics completely. An alternative is to separate desired subgraphs and use single network algorithms which defeats the purpose of modeling as attribute graphs and is inefficient.

\subsection{The  Decoupling Approach}

\begin{figure}[h]
\centering
\vspace{-10pt}
\includegraphics[width=\columnwidth]{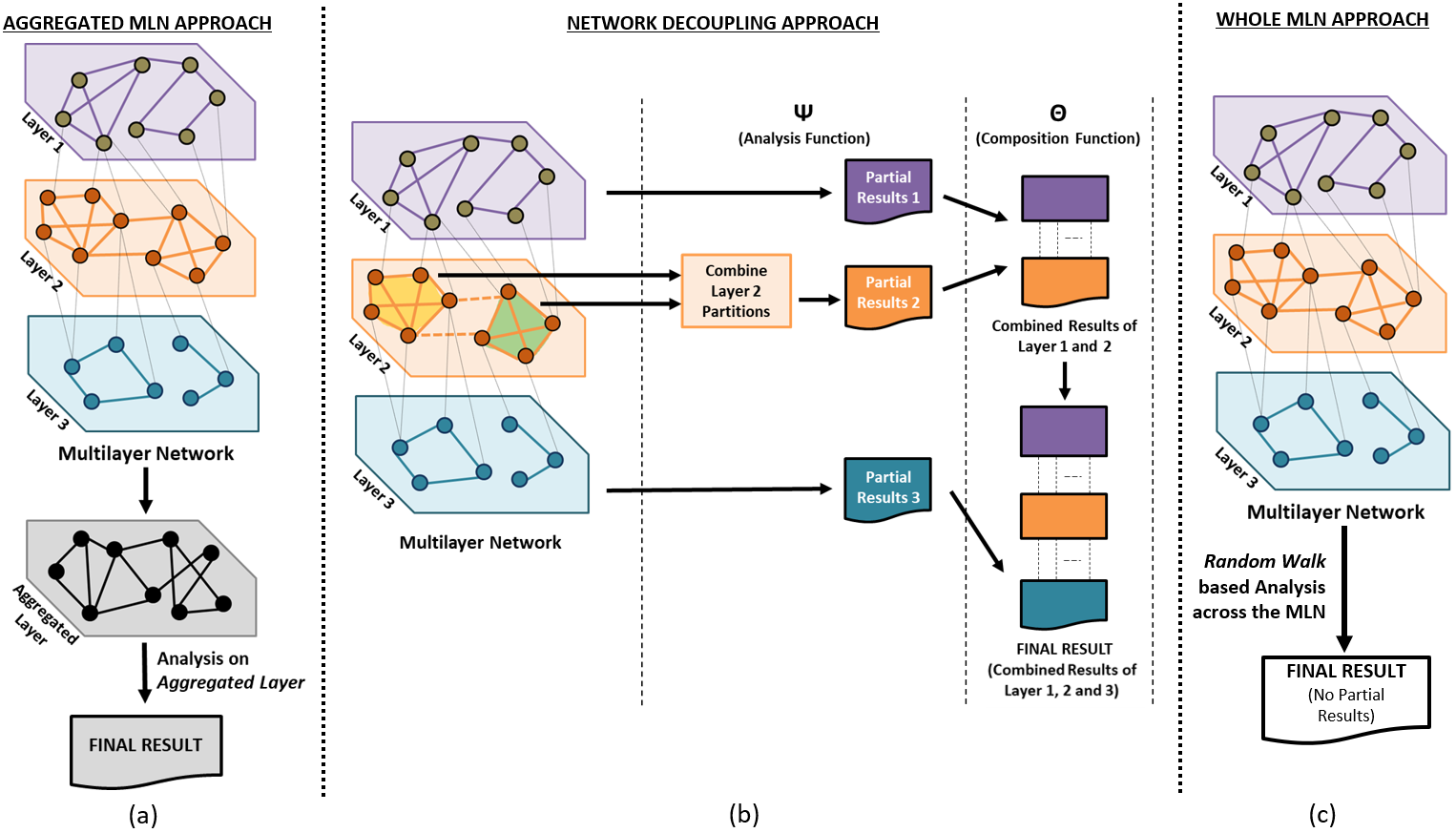}
%\vspace{-20pt}
\caption{Alternative Approaches for MLN Analysis}
\vspace{-10pt}
\label{fig:MLNalts}
\end{figure}

Network decoupling is a method by which MLNs can be analyzed \textit{without being transformed}. The decoupling approach preserves the structure and semantics of the layers natively in the result and at the same time can take advantage of the existing algorithms. The {\em network decoupling}  approach developed in ~\cite{ICCS/SantraBC17,ICDMW/SantraBC17} is the equivalent of ``divide and conquer" for MLNs. This is summarized in Figure~\ref{fig:MLNalts}(b) and is applied as follows, for a given analysis function $\Psi$ and composition function, $\Theta$:
 \begin{packed_itemize}
 \item {\bf(i)} Use the {analysis function $\Psi$} to analyze each layer individually.
 
 \item {\bf(ii)} Second, for any two chosen layers, apply a \textit{composition function $\Theta$} to compose the partial results from each layer to generate intermediate results.
 
 \item {\bf(iii)} Finally, apply the composition process until the expression is computed.
 \end{packed_itemize}
 
 This is in contrast to current approaches described earlier. Figure~\ref{fig:MLNalts}(a) indicates aggregation-based approaches where structure and semantics are lost (sans mapping.) Figure~\ref{fig:MLNalts}(c) illustrates MLN approaches where only inter-layer edges are used instead of all edges.
 
\underline{\em Advantages:} The decoupling approach has several advantages over the traditional methods. By using the aggregation approach, information pertaining to the individual layers is lost and it is difficult to measure their relative importance to the system as a whole.  In contrast, network decoupling retains the semantic information of each layer and therefore their individual importance and contribution can be measured. The ``divide and conquer" approach also facilitates the mix and match of the features and relationships. In the aggregation approach, each time a subset of features is selected, the analysis has to be recomputed, even when the subsets might have overlaps. This leads to redundant computations. Using the decoupling approach, redundant analysis are avoided, since each layer, corresponding to a particular feature is analyzed separately, and then combined. Finally and importantly, the structure and semantics are preserved in the results explicitly a there is no conversion needed in this approach.

\underline{\em Challenges.}  The decoupling approach can be applied to both HoMLN and HeMLN and hence to HyMLN as well. Moreover, the success of this approach is dependent on correctly matching the analysis objectives using appropriate $\Psi$ as the aggregate function and  $\Theta$ as the composition function. In Section~\ref{sec:expression-generation}, we show how the network decoupling approach can be applied for our data sets and appropriately determine $\Psi$ and $\Theta$ for the diverse analysis objectives~\ref{analysis:Airline1} through ~\ref{analysis:Covid1}.
 
\begin{comment}
\begin{figure}
\centering
\vspace{-40pt}
\includegraphics[width=\columnwidth]{figures/decoupling-v3.jpg}
\caption{Network Decoupling for MLN Analysis}
\vspace{-10pt}
\label{fig:decoupling}
\end{figure}
\end{comment}

A number of algorithms that use the decoupling approach have been developed for community detection for both HoMLN~\cite{ICCS/SantraBC17} and HeMLN~\cite{Arxiv/SantraKBC20} as well as centrality detection~\cite{ICDMW/SantraBC17} for HoMLN. An algorithm for substructure discovery on MLNs has been developed in~\cite{msThesis/Rai20}. There are also some algorithms that compute substructures ~\cite{Boden:2012:MCS:2339530.2339726} and community~\cite{magnani2019community} directly on MLNs without collapsing or aggregating them. In this paper, we use the decoupling based algorithms as they cover the needs of all analysis objectives under consideration.

\section{Generating Analysis Expressions From Objectives}
\label{sec:expression-generation}

As indicated in Section~\ref{sec:eer-to-mln}, the modeling of data sets and the generation of a MLN (HoMLN/HeMLN/HyMLN) depends mainly on the relationships identified on the entities in the data set. Typically, self-relationships generate HoMLNs, n-ary (mostly binary as we do not use/support hyper edges in MLNs yet) relationships generate HeMLNs. The EER diagram is also converted to a relational schema for drill-down and additional processing as specified in objectives. The details of creating edges in each layer come from the attributes (deemed as  parameters) from the relationships used for MLN generation.  For example, for~\ref{analysis:DBLPHeMLN2}, the 3-year period explicitly provided in the analysis objective is modeled in the EER diagram as an attribute of the relationship same-interval and is used for creating the layer graph. Each three consecutive years form a clique in that layer. If a threshold is needed for creating edges of the layer \textit{Year} to capture similarity, it becomes a parameter of the analysis alternatives. As an example, for \ref{analysis:DBLPHeMLN1} 3 publications together has been used for the parameter \textit{num-of-papers} attribute associated with the collaborates-with relationship that generates the \textit{Author} layer. This is a parameter that can be modified to perform a different set of analysis. The MLN model nor the expressions change, except the graph of the layer generated based on this parameter value. 

In addition to these, table~\ref{table:ThetaLookUp-DBLP} is generated indicating possible $\Theta$ for each pair of layers for the DBLP data set. Similar tables are generated for each data set. This is dependent on the outcome of modeling. Table~\ref{table:PsiLookUp-DBLP} is a table that indicates available $\Psi$ options for each layer. A similar table is generated for  each layer produced during modeling of each data set. This depends on algorithms available for computing expressions as shown in Figure~\ref{fig:life-cycle-flow}. This is independent of the modeling step.

For this paper, we assume community and centrality (both node and closeness) for $\Psi$. Other possible $\Psi$ options (shown in Table \ref{table:PsiLookUp-DBLP}) can be any graph-based analysis approach like degree centrality, interesting substructure discovery and so on. The complement of a graph (unary NOT) is another $\Psi$ option that produces a graph with complement set of edges based on the input graph. For composition of homogeneous layers, we assume binary AND and OR binary compositions.% and unary NOT composition for any individual layer.
For composing heterogeneous layers, we assume Maximum Weighted Matching (MWM) bipartite approach~\cite{edmonds1965maximum} as discussed in~\cite{Arxiv/SantraKBC20}.

Briefly, algorithms for computing 2-layer communities for HoMLNs use either Boolean AND or OR composition. Algorithms for these are in~\cite{ICCS/SantraBC17,ICDMW/SantraBC17}. Boolean NOT operator (complement of a graph) can be used for any layer and further composed with other layers using AND or OR. Multiple layer community computation is done by applying the operators on the result of the previous step. The order of Boolean operator application can be user-specified (or generated as we show in this paper.)

Composing HeMLNs for community detection is challenging since the entities are of different types. As described in~\cite{Arxiv/SantraKBC20}, each community is considered to be a meta-node. Two meta-nodes in two different layers are connected if there is at least one inter-layer edge between them. The weight of these edges (meta-edges) between the meta-nodes is given by the number of inter-layer edges between them. This construction creates a bipartite graph. These meta nodes (communities) in the bipartite graph are paired using the composition function ($\Theta$) {Maximum Weighted Matching (MWM) as proposed by Jack Edmonds}~\cite{edmonds1965maximum}. Thus, the \textit{paired meta-nodes} correspond to the heterogeneous MLN communities.

\begin{figure}[h]
\centering
\vspace{-15pt}
\includegraphics[width=\columnwidth]{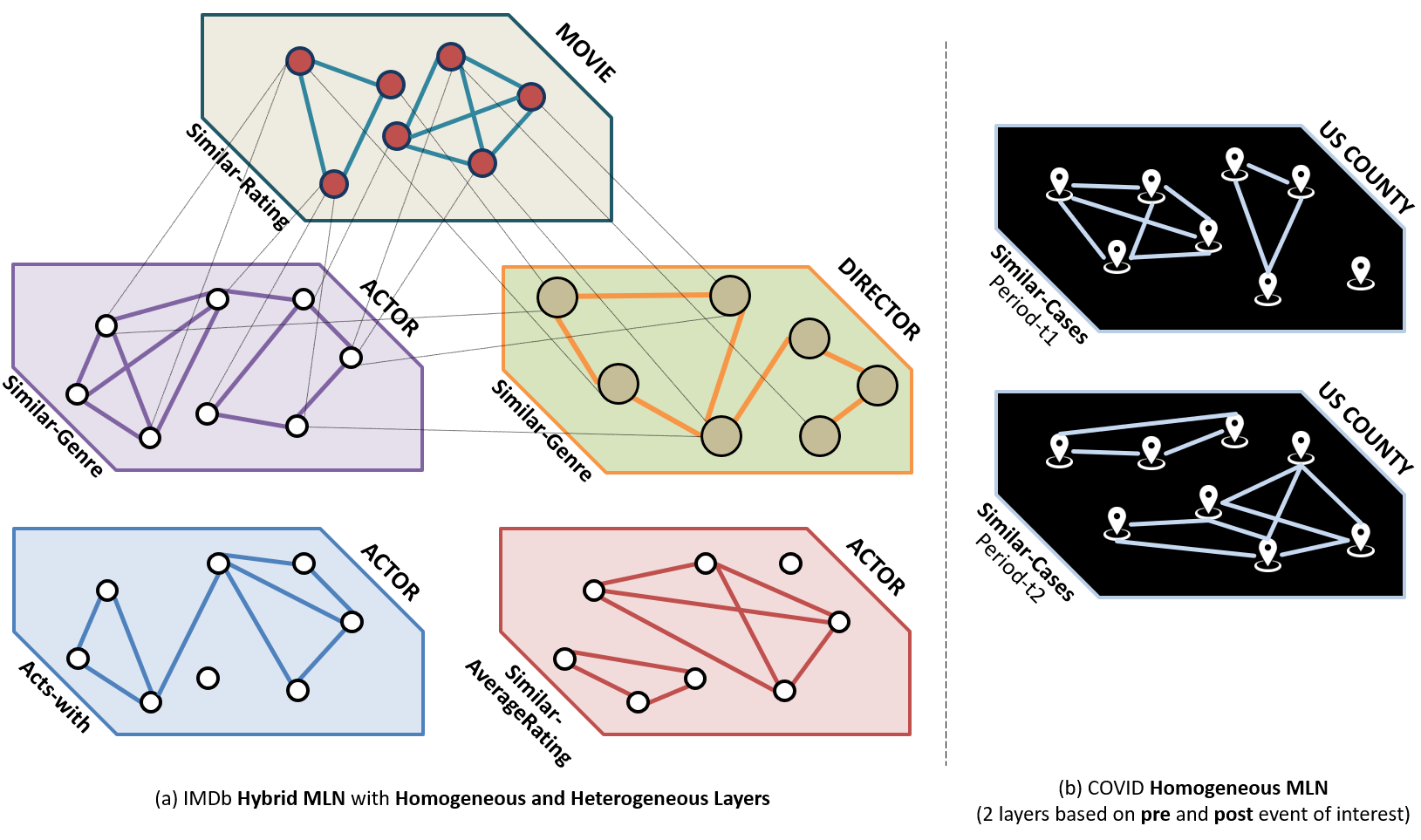}
\vspace{-10pt}
\caption{Multilayer Networks Generated for remaining Analysis Objectives}
\vspace{-20pt}
\label{fig:otherMLNs}
\end{figure}

The MLNs derived for the US-Airlines and the DBLP data sets (described in Section~\ref{sec:analysis-life-cycle}) are shown in Figures~\ref{fig:USAirline-Model} and ~\ref{fig:DBLP-MLN}, respectively. In Figures~\ref{fig:otherMLNs}, we show the MLNs derived after modeling the IMDb (Figure ~\ref{fig:otherMLNs} (a)) and Covid (Figure ~\ref{fig:otherMLNs} (b)) data sets as EER diagrams and applying the algorithm in ~\cite{ER/KomarSBC20}. The figures do not include a few inter-layer edges for IMDb MLN to maintain clarity. The EER diagrams and other relations derived for drill-down follow the approach illustrated for the other two data sets.

We use the characteristics of the MLNs generated to create a table for each data set for looking up $\Theta$ during translation. Another table is created for each layer in each data set to indicate what operations ($\Psi$) are available. The $\Psi$ (Table \ref{table:PsiLookUp-DBLP}) and $\Theta$ (Table \ref{table:ThetaLookUp-DBLP}) lookup tables have been shown for DBLP MLN. In addition, as the current approach is based on extracted keywords and their interpretation (e.g., nouns as layers, verbs as $\Psi$, and conjunctions as $\Theta$), another table is used for lookup during translation. This is shown in Table ~\ref{table:keyword-map} which groups keywords and their possible synonyms in each category with the corresponding choice for computation. The scope of translation depends on the available $\Psi$ and $\Theta$. Objectives that cannot be mapped to available computable operations are indicated as such. Then, we convert each analysis objective for that data set to expressions that can be computed on the MLN model. That is $\Psi$ and $\Theta$ are inferred. Also, additional computations on the result (e.g., sorting, ranking, top-k, ...) may need to be  performed based on the wording in the objectives in addition to translation.

\begin{table}[h]
\renewcommand{\arraystretch}{1.3}
%%\begin{table}[h]
\centering
\scriptsize
%\vspace{-20pt}
        \begin{tabular}{|m{1cm}|m{1.34cm}|m{2.2cm}|m{1.5cm}|m{1.7cm}|m{1.5cm}|}
        %\begin{tabular}{|m{0.5cm}|m{2.8cm}|m{2.6cm}|m{0.8cm}|}
        \hline
        \textbf{L1} & Community & Degree Centrality & Closeness Centrality & Substructure Discovery & Complement (NOT)\\
        \hline 
        \textbf{L2} & Community & Degree Centrality & Closeness Centrality & Substructure Discovery & Complement (NOT) \\
        \hline
        ...         &           &                   &                       &                       &                   \\
        \hline
        %%\hline
        %%\textbf{L3} & Community & Degree Centrality & Closeness Centrality & Substructure Discovery & Complement (NOT) \\
        %%\hline
        %%\textbf{L4} & Community & Degree Centrality & Closeness Centrality & Substructure Discovery & Complement (NOT) \\
        %%\hline
        %%\textbf{L5} & Community & Degree Centrality & Closeness Centrality & Substructure Discovery & Complement (NOT) \\
        %%\hline
        %%\textbf{L6} & Community & Degree Centrality & Closeness Centrality & Substructure Discovery & Complement (NOT) \\
        %%\hline
        \textbf{L7}  & Community & Degree Centrality & Closeness Centrality & Substructure Discovery & Complement (NOT) \\
        \hline
        \hline
        \textbf{Layer IDs} & \multicolumn{5}{m{10cm}|}{ \textbf{L1:}AUTHOR-Collaborates-With, \textbf{L2:}AUTHOR-Collaborates-in-VLDB, \textbf{L3:}AUTHOR-Collaborates-in-SIGMOD, \textbf{L4:}AUTHOR-Collaborates-in-DASFAA, \textbf{L5:}AUTHOR-Collaborates-in-DaWaK, \textbf{L6:}PAPER-Same-Conference, \textbf{L7:}YEAR-Same-Interval} \\
        \hline
        
\end{tabular}
%\vspace{-10pt}
\caption{$\Psi$ Lookup Table for DBLP MLN Layers}
\vspace{-10pt}
\label{table:PsiLookUp-DBLP}
%%\end{table}
\end{table}

\begin{table}[h]
\renewcommand{\arraystretch}{1.3}
%%\begin{table}[h]
\centering
\scriptsize
\vspace{-20pt}
        \begin{tabular}{|*{8}{c|}}
                                                                                \cline{1-2}
        \textbf{L1} &                                                      \\  \cline{1-3}
        \textbf{L2} & AND, OR &                                           \\  \cline{1-4}
        \textbf{L3} & AND, OR & AND, OR &                                \\  \cline{1-5}
        \textbf{L4} & AND, OR & AND, OR & AND, OR &                        \\  \cline{1-6}
        \textbf{L5} & AND, OR & AND, OR & AND, OR & AND, OR &              \\  \cline{1-7}
        \textbf{L6} & MWM & MWM & MWM & MWM & MWM &     \\ 
        \cline{1-8}
        \textbf{L7} & MWM & MWM & MWM & MWM & MWM & MWM & \\ 
        \cline{1-8}
        
        \hline
        & \textbf{L1} & \textbf{L2} & \textbf{L3} & \textbf{L4} & \textbf{L5} & \textbf{L6} & \textbf{L7} \\
        \hline
        \hline
        \textbf{Layer IDs} & \multicolumn{7}{m{10cm}|}{ \textbf{L1:}AUTHOR-Collaborates-With, \textbf{L2:}AUTHOR-Collaborates-in-VLDB, \textbf{L3:}AUTHOR-Collaborates-in-SIGMOD, \textbf{L4:}AUTHOR-Collaborates-in-DASFAA, \textbf{L5:}AUTHOR-Collaborates-in-DaWaK, \textbf{L6:}PAPER-Same-Conference, \textbf{L7:}YEAR-Same-Interval} \\
        \hline
\end{tabular}
%\vspace{-10pt}
\caption{$\Theta$ Lookup Table for DBLP MLN Layer Pairs}
\vspace{-20pt}
\label{table:ThetaLookUp-DBLP}
%%\end{table}
\end{table}

The challenge here is the automation of analysis expression generation given objectives in English that is meant for human consumption. From our experience in analyzing many data sets for diverse objectives, these objectives can be expressed in multiple ways and can be manually translated into several alternative expressions. This is due to multiple interpretations which lead to multiple expressions for the same objective. Figuring out the order of computations of expressions is another challenge. As explained earlier, we use a keyword-based heuristics approach in this paper as there is no general purpose NLP-based approach that we are aware of. Below, we explain our approach and provide intuitive explanations of how they are generated.

Similar to the heuristics used for EER modeling, nouns, verbs, and conjunctions are identified from objectives. We have highlighted parts of each objective used for expression generation. Phrases in the objective are \underline{underlined}, \textit{italicized}, and shown in \textbf{bold}  below to indicate their use, respectively, for layer selection (using nouns), analysis function ($\Psi$) determination (using verb forms), and composition function ($\Theta$) identification (using conjunctions) for the generation of the expression. These can be isolated by an NLP keyword/phrase analysis and used for looking up $\Psi$ and $\Theta$ using the three tables indicated earlier. If a mapping is not possible (or keyword not properly recognized), lookup of the tables will fail. Additional computations, as needed, are also inferred from the keywords/phrases in the objective, in the form of \texttt{FILTER} operation. For example, predict is translated to ranking based on sort and choosing the top-k values. Listing of top k entities translates to sorting and retaining top 5 entities. Here, for the list of the objectives, the \texttt{FILTER} operation is applied (as per requirement) at the end of the analysis expression or earlier as shown with each expression.

\begin{table}[h]
    \centering
    \begin{tabular}{|m{4.8cm}|m{4cm}|}
    \hline
    \textbf{Keywords/Phrases} & \textbf{Mapped to $\Psi$/$\Theta$} \\
    \hline
    \textit{group}, \textit{cluster}, \textit{strong/dense group} & $\Psi$ = Community \\
    \hline
    \textit{coverage} & $\Psi$ = Closeness Centrality \\
    \hline
    \textit{direct neighbors, hubs} & $\Psi$ = Degree Centrality \\
    \hline
    \textit{frequent/interesting patterns} & $\Psi$ = Substructure Discovery \\
    \hline
    \textit{never}, \textit{not} & $\Psi$ = Complement (NOT) \\
    \hline
%    Conjunctions = \textbf{and}, \textbf{but}, \textbf{yet}, \textbf{who} & $\Theta$ = AND for HoMLN layers\\
    \textbf{and}, \textbf{but}, \textbf{yet} & $\Theta$ = AND for HoMLN layers\\
    \hline
    \textbf{or}, \textbf{either} & $\Theta$ = OR for HoMLN layers\\
    \hline
%    Conjunctions = \textbf{who}, \textbf{when}, \textbf{and}, \textbf{but}, \textbf{yet}; Adverbs = \textbf{for each}, \textbf{-wise} & $\Theta$ = MWM for HeMLN layers\\
    \textbf{and}, \textbf{but}, \textbf{yet}, \textbf{for each, for every} & $\Theta$ = MWM for HeMLN layers\\
    \hline

    \end{tabular}
    \caption{Keyword-based Lookup for Operators}
    \vspace{-20pt}
    \label{table:keyword-map}
\end{table}

\subsection{Expressions Generated For US-Airlines Objectives} 
  \begin{enumerate}  [label={\textbf{(A\arabic*)}}]  
    \item
    For \underline{American}, \underline{Southwest}, \underline{Spirit}, \underline{Delta}, \underline{Frontier} and \underline{Allegiant} Airlines rank the top five cities, that  provide  the maximum  \textit{coverage}.
\vspace{5pt}

Coverage in \ref{analysis:Airline1} corresponds to cities from which \textit{one can cover most number of cities using least number of flights}. Hence, \textit{Coverage} can be used to translate the objective intent to \textit{closeness centrality} as $\Psi$ as shown in Table~\ref{table:keyword-map}. For airlines, since low closeness centrality value means low average distance (number of flights) to cover cities, that will provide cities which can be ranked further to fetch the top-5. The expression derived is shown below.

    \texttt{Expression: $\Psi$(Each layer), FILTER = top-5 using closeness value; \\$\Psi$ = Closeness Centrality; $\Theta$ = N/A}

\vspace{5pt}

\item Predict which city (taking its population into consideration) could be selected as the next hub(s) for \underline{Allegiant} Airlines to expand  its \textit{coverage} and avoid competition with \underline{other airlines}. 
\vspace{5pt}

Analysis \ref{analysis:Airline3} is more complicated and perhaps beyond the current approach. However, it is clear from the objective and the mapping table that $\Psi$ is closeness centrality. However, prediction is for cities that are not currently Allegiant hubs which requires eliminating the hubs generated from the $\Psi$ computation. Avoiding competition indicates non-overlap with cities that have high closeness value for other airlines resulting in the expression shown below. $\Psi$ and $\Theta$ can be looked up. Sorting on population at the end is clear. Using this simple approach, it is difficult to derive the differences. This is a good example of the challenge we mentioned earlier that is difficult for the keyword-based heuristics translation!

%%since it requires prediction/identification of future hubs. Note that this objective also requires use of city relation attributes beyond aggregate analysis. Prediction indicates selecting the best city which translates to ranking cities. First, we have to select cities that are not hubs already for the airline under consideration. For this, we need to compute hubs and retain non-hubs. Then, we need to eliminate the cities in the non-hub list with hubs of the other airline to avoid \textit{competition}. Here too the analysis objective keyword indicates closeness centrality for the use of $\Psi$. As this objective involves comparisons with other airlines hubs, $\Theta$ comes into picture. For \ref{analysis:Airline3}, we match the \textit{set of non-hubs that have high closeness centrality} of the target airline (Allegiant in our case) to be expanded against \textit{the set of hubs of the competing airlines}. Here we infer AND operation (as $\Theta$) on the hub sets of competing airlines to find the common hubs that are discarded as they are already hubs for the competing airlines. Finally, the resulting cities are ranked based on population to get the final result.

Note that this objective can be specified and computed for any airline. In addition, other city attributes (than population), such as mean/median income or combinations can be used. These do not affect either the model or the expression generation. Ordering the resulting cities based on the population (or any attribute) and choosing the top one is the last step.
\vspace{5pt}

\begin{comment}

\texttt{Expression: \{{Allegiant cities} - $\Psi$(Allegiant-Direct-Flight hubs)\} -  $\Psi$(Each of Remaining Airlines hubs Separately), \\ Filter = sort-by-population; \\ $\Psi$ = Closeness Centrality; $\Theta$ = AND}

\end{comment}

\texttt{Expression: ($\Psi$(USCITY-Allegiant-Direct-Flight) - \\ \{$\Psi$(USCITY-Allegiant-Direct-Flight) $\Theta$  $\Psi$(Each of Remaining Airlines Separately)\}) - \{Active Allegiant Hubs\}; FILTER = sort-by-population; \\ $\Psi$ = Closeness Centrality; $\Theta$ = AND}

\end{enumerate}

\subsection{Expressions Generated for DBLP Objectives} 

\begin{enumerate} [label={\textbf{(A\arabic*)}},resume]  

%\item \textbf{For each} disjoint 3-\underline{year} \textit{interval}, find the \textit{strong} co-\underline{author} \textit{group}s \textbf{who} were most actively publishing.
\item \textbf{For each} 3-\underline{year interval} \textit{group}, find the most actively publishing \textit{strong} \underline{author collaboration} \textit{group}s.

Once the layers are identified for \ref{analysis:DBLPHeMLN1} as AUTHOR-Collaborates-With and YEAR-Same-Interval, it can be looked up from table \ref{table:ThetaLookUp-DBLP} as a HeMLN operation. Based on the keyword mapping Table \ref{table:keyword-map}, the required analysis is community detection (phrases = \textit{group, strong group}), which have to be combined using MWM composition due to the identification of heterogeneous layers (phrase = \textbf{for each} serves as conjunction), resulting in the expression shown below. %The phrases \textit{strong groups}, \textit{3-year interval} and \textbf{most actively publishing} indicate community detection in individual and across layers, resulting in the expression shown below.

\texttt{Expression: $\Psi$(YEAR-Same-Interval) \textbf{$\Theta$} $\Psi$(AUTHOR-Collaborates-With); \\ $\Psi$ = Community; $\Theta$ = MWM}
    
    \vspace{5pt}
    
%\sharma{4/7/21}{requires some discussion/change}
    
\item
    %For the \textbf{most popular} \underline{\textit{collaborators}} in each \underline{\textit{conference}}, identify the \textit{3-year \underline{period(s)}} when they were \textbf{most active}.
%    \underline{Conference}\textbf{-wise} \textbf{for each} of the most popular \underline{collaboration} \textit{group}, identify the 3-\underline{year} \textit{interval}(s) \textbf{when} they were most active.

    \textbf{For each} \underline{conference-based paper} \textit{group}, find the most popular \underline{author} \underline{collaboration} \textit{group} and further \textbf{for each} of them identify their most active 3-\underline{year interval} \textit{group}(s).

%    \sharma{4/4/21}{not sure why the expression is paper theta au-collab theta year? the MLN does not even show inter-layer edges between paper and author! need to discuss}

    For \ref{analysis:DBLPHeMLN2}, 3 layers are identified from the objective. The italicized phrases indicate community detection within layers (phrases = \textit{group}.). Further,  MWM is used as the $\Theta$ throughout to generate the final HeMLN communities (phrases = \textbf{for each}, \textbf{and}), based on mapping and lookup tables \ref{table:keyword-map} and \ref{table:ThetaLookUp-DBLP}. Order is implied by the keyword further.

    \texttt{Expression: ($\Psi$(PAPER-Same-Conference) \textbf{$\Theta_1$} \\ $\Psi$(AUTHOR-Collaborates-With)) \textbf{$\Theta_2$} $\Psi$(YEAR-Same-Interval); \\ $\Psi$ = Community; $\Theta_i$ = MWM}
\vspace{5pt}

\item Identify \underline{author collaboration} \textit{group}s who have published in conferences \underline{VLDB} \textbf{and} \underline{SIGMOD}, \textbf{but} have \textit{never} published in conferences \underline{DASFAA} \textbf{and} \underline{DaWaK}.

%    \sharma{4/4/21}{let us add parenthesis in the expressions to show order of computation. this is just to indicate that the order comes out of this process}
    Here 4 layers are identified. The objective clearly indicates the need for the NOT operator (phrase=\textit{never} for the layers Author-Collaborates-in-DASFAA and Author-Collaborates-in-DaWaK). The community indication is clear (phrase = \textit{group}). Moreover, all are homogeneous layers which support boolean composition, as looked up from table \ref{table:ThetaLookUp-DBLP}. First, we compute 2-layer community using AND (between Author-Collaborates-in-VLDB and Author-Collaborates-in-SIGMOD), which is again AND composed with the NOT communities of the other two conferences (phrases = \textbf{and}, \textbf{but}.) Thus, generating the final expression.
    \vspace{5pt}
    
    \texttt{Expression: ($\Psi$(AUTHOR-Collaborates-in-VLDB) $\Theta_{1}$ \\ $\Psi$(AUTHOR-Collaborates-in-SIGMOD)) $\Theta_3$ \\ ($\Psi$(NOT(AUTHOR-Collaborates-in-DASFAA)) $\Theta_2$ \\ $\Psi$(NOT(AUTHOR-Collaborates-in-DaWaK))); \\ $\Psi$ = Community; $\Theta_i$ = AND}
    \vspace{5pt}
    
    Note that the expression generated for this can be re-written to improve computation efficiency. We do not further discuss that in this paper except to indicate that expressions generated can be further optimized using any means available without affecting correctness.
    \end{enumerate}

\subsection{Expressions Generated for IMDb Objectives} 

  \begin{enumerate}  [label={\textbf{(A\arabic*)}}, resume]  
    
  \item \label{analysis:IMDbHoMLN1} %\textit{Cluster groups}  of  \underline{co-actors} who  \textbf{have  also worked}  in  movies  with \underline{high  ratings}.
    %%%ALTERNATIVE: 
%    \textit{Cluster}  \underline{co-actors} \textbf{who} have a \underline{similar average rating}.

    \textit{Cluster}  \underline{actors} who have \underline{acted together} \textbf{and} have a \underline{similar average rating}.
\vspace{5pt}

Expression generation for  \ref{analysis:IMDbHoMLN1} is quite straightforward with the identification of ACTOR-Acts-with and ACTOR-Similar-AverageRating layers. The \textit{cluster} keyword indicates community computation on the individual layers that are combined further using the AND composition (conjunction = \textbf{and}.)

\texttt{Expression: $\Psi$(ACTOR-Acts-with) $\Theta$ $\Psi$(ACTOR-Similar-AverageRating); $\Psi$ = Community; $\Theta$ = AND}

    \vspace{5pt}

\item \label{analysis:IMDbHoMLN2} Find the \textit{group}s of \underline{actors} who have \textit{never} \underline{acted together}, \textbf{but} are highly \underline{rated on an average} \textbf{and} have worked in \underline{similar genres}.

%\sharma{4/4/21}{again, expression order DOES NOT follow objective order! let us discuss why}

For \ref{analysis:IMDbHoMLN2}, the layers identified are ACTOR-Similar-AverageRating, ACTOR-Similar-Genre, and ACTOR-Acts-with. These are homogeneous layers as well and community detection is required (phrase = \textit{group}.) Based on the objective, NOT is applied to layer ACTOR-Acts-With (phrase = \textit{never}) before composing it with layer Actor-Similar-AverageRating, which is then finally composed with layer Actor-Similar-Genre using AND composition (phrases = \textbf{but}, \textbf{and}) as $\Theta$s throughout leading to the generated expression. sort-on-AverageRating needs to be inferred from the keyword highly in the objective to output top k results. Order of composition is inferred from the objective as given.

\texttt{Expression: ($\Psi$(NOT(ACTOR-Acts-with)) $\Theta_1$ \\ $\Psi$(ACTOR-Similar-AverageRating)) $\Theta_2$ $\Psi$(ACTOR-Similar-Genre), \\ FILTER = sort-on-AverageRating; \\ $\Psi$ = Community; $\Theta_i$ = AND}
    \vspace{5pt}

%\item \label{analysis:IMDbHeMLN1} Identify \underline{genre-based} \textit{group}s of  \underline{actors}  \textbf{and}  \underline{directors}  \textbf{who}  have  close collaborations.
\item \label{analysis:IMDbHeMLN1} Identify \underline{genre-based} \textit{group}s of  \underline{actors}  \textbf{and}  \underline{directors}  having  strong collaborations.

This objective is quite straightforward and is similar to others done earlier for 2 layer HeMLN community expression generation.

\texttt{Expression: $\Psi$(ACTOR-Similar-Genre) $\Theta$ $\Psi$(DIRECTOR-Similar-Genre); \\ $\Psi$ = Community; $\Theta$ = MWM}

    \vspace{5pt}

\item \label{analysis:IMDbHeMLN2} %Identify, for each \underline{movie rating}, the \textbf{most popular} \underline{genre-based actor} and \underline{director} \textit{groups} who have \textbf{strong collaborations}.

%Identify, \textbf{for each} \underline{movie rating} \textit{range} the \underline{genre-based} most popular \underline{actor} \textbf{and} most popular \underline{director} \textit{group}s. From this result, find the \underline{actor} \textbf{and} \underline{director} \textit{group}s \textbf{who} have strong collaborations.

Identify, \textbf{for each} \underline{movie rating} \textit{group} the \underline{genre-based} most popular \underline{actor} \textbf{and} most popular \underline{director} \textit{group}s. From this result, find the \underline{actor} \textbf{and} \underline{director} \textit{group}s having strong collaborations.

%Identify, the \textbf{most popular} \underline{genre-based actor} \textit{groups} for each \underline{movie rating}, and also the \textbf{most popular} \underline{director} \textit{groups}. From these results, find the \underline{actor} and \underline{director} \textit{groups} who have \textbf{strong collaborations}.
\vspace{5pt}

Here MOVIE-Similar-Rating, ACTOR-Similar-Genre and DIRECTOR-Similar-Genre are the 3 identified layers which needs to be processed for detecting the HeMLN communities ($\Psi$ phrases = \textit{group}, \textit{range}; $\Theta$ phrases = \textbf{for each}, \textbf{and}.) Based on the objectives, the layer MOVIE-Similar-Rating becomes a common layer that is composed with ACTOR-Similar-Genre and DIRECTOR-Similar-Genre layers. This ordering is derived from the objective and translated to the generated expression by the subscripts of $\Theta$ - $\Theta_{M,A}$ for MOVIE-Similar-Rating and ACTOR-Similar-Genre composition, and $\Theta_{M,D}$ for MOVIE-Similar-Rating and DIRECTOR-Similar-Genre composition. Finally, the DIRECTOR-Similar-Genre and ACTOR-Similar-Genre are composed. MWM (Maximum Weighted Matching) is used for the $\Theta$ operation, thus completing the generated expression.

\texttt{Expression: (($\Psi$(MOVIE-Similar-Rating) $\Theta_{M,A}$ $\Psi$(ACTOR-Similar-Genre)) $\Theta_{M,D}$  $\Psi$(DIRECTOR-Similar-Genre)) $\Theta_{D,A}$ $\Psi$(ACTOR-Similar-Genre); \\ $\Psi$ = Community; $\Theta$ = MWM}

\end{enumerate}

%%%\noindent \textbf{DBLP HoMLN Analysis:} For \ref{analysis:DBLPHoMLN1}, we need to focus on the co-author groups, thus analysis function ($\Psi$) corresponds to community detection. For co-author groups in high ranked conference, we need to detect communities in the VLDB and SIGMOD layer, individually. For co-author groups \textit{not publishing} in low to medium ranked conferences, we need to first apply the NOT operation on the DASFAA and DaWaK layers followed by community detection. Finally, the application of the Boolean AND composition function on the 4 four sets of the resulting layer-wise communities will fulfill the objective.  

%%%\noindent {{\bf  IMDb and DBLP HeMLN Analysis:}} For the objectives related to these MLNs we have to find the communities comprising of different types of entities. More specifically, community of actors and directors for \ref{analysis:IMDbHeMLN1}, communities of actors, directors and movies for \ref{analysis:IMDbHeMLN2}, communities of authors and papers for \ref{analysis:DBLPHeMLN1} and communities author, papers and years for \ref{analysis:DBLPHeMLN2}. %Thus the analysis function is community detection for heterogeneous MLNs. %Since we had already computed the communities of actors and directors as part of {\bf(A6)}, we need not recompute them in {\bf (A7)}.

\subsection{Expressions Generated for Covid Objectives}

  \begin{enumerate}  [label={\textbf{(A\arabic*)}}, resume]  
  
      \item \label{analysis:Covid1} Visualize how the  geographical regions corresponding to the \textit{cluster}s of \underline{US counties} with rise (or decline) in  
       \underline{daily confirmed cases} shift in the - \\ 
    i) \underline{consecutive 7-day periods} \underline{pre and post 2020 spring break} and, \\
    ii)  \underline{month-apart 3-day periods} \underline{pre and post vaccination drive}?  

For the spring break and vaccination based identified layers in \ref{analysis:Covid1} (i) and \ref{analysis:Covid1} (ii), respectively, community detection (phrase = \textit{group}) needs to be performed, as captured by the final expression. 

\texttt{Expression: $\Psi$(Each Layer); $\Psi$ = Community; $\Theta$ = N/A}

\end{enumerate}

To summarize, Table \ref{table:computation} gives the mapping of each analysis question \ref{analysis:Airline1} to \ref{analysis:Covid1} to their actual computation specification (in \textit{left} to \textit{right} order), analysis function ($\Psi$) and composition function ($\Theta$).

\begin{table}[h]
\renewcommand{\arraystretch}{1.6}
%%\begin{table}[h]
\centering
%\vspace{-20pt}
        \begin{tabular}{|m{1.2cm}|m{6.3cm}|m{1.4cm}|m{1.3cm}|}
        %\begin{tabular}{|m{0.5cm}|m{2.8cm}|m{2.6cm}|m{0.8cm}|}
            \hline
            \multirow{2}{0.5cm}{\textbf{Analysis}} & \multicolumn{3}{c|}{\textbf{Mapping}} \\
            \cline{2-4}
            & \multicolumn{1}{c|}{\textbf{Computation Expression}} &  \multicolumn{1}{c|}{\textbf{$\Psi$}} & \multicolumn{1}{c|}{\textbf{$\Theta_i$}}  \\
            \hline
            
            \hline
            \hline
            \rowcolor{orange!40}\multicolumn{4}{|p{11.5cm}|}{\textbf{US-Airline {HoMLN} Layers:} USCITY-American-DirectFlight, USCITY-Southwest-DirectFlight, USCITY-Delta-DirectFlight, USCITY-Spirit-DirectFlight, USCITY-Frontier-DirectFlight, USCITY-Allegiant-DirectFlight} \\
            \hline
            \ref{analysis:Airline1} & Apply $\Psi$ on each layer, FILTER = top-5 using closeness value & Closeness Centrality & N/A \\
           \hline
            %\ref{analysis:Airline2} & \textit{p} major airline layers; \textit{q} minor airline layers & Hub (degree) & $>$ \\
            %\hline
            \ref{analysis:Airline3} & ($\Psi$(USCITY-Allegiant-Direct-Flight) - \{$\Psi$(USCITY-Allegiant-Direct-Flight) $\Theta$  $\Psi$(Each of Remaining Airlines Separately)\}) - \{Active Allegiant Hubs\}; FILTER = sort-by-population & Closeness Centrality & AND \\
            \hline
        
            \hline     
            \hline
        \rowcolor{green!30}\multicolumn{4}{|p{11.5cm}|}{\textbf{DBLP MLN Layers}: AUTHOR-Collaborates-With, AUTHOR-Collaborates-in-VLDB, AUTHOR-Collaborates-in-SIGMOD, AUTHOR-Collaborates-in-DASFAA, AUTHOR-Collaborates-in-DaWaK, PAPER-Same-Conference, YEAR-Same-Interval} \\            
            \hline
            \ref{analysis:DBLPHeMLN1} & YEAR-Same-Interval \textbf{$\Theta$} AUTHOR-Collaborates-With & Community  & MWM \\
            \hline
            \ref{analysis:DBLPHeMLN2} & PAPER-Same-Conference \textbf{$\Theta_1$} AUTHOR-Collaborates-With \textbf{$\Theta_2$} YEAR-Same-Interval & Community & MWM\\
            \hline           \ref{analysis:DBLPHoMLN1} & (AUTHOR-Collaborates-in-VLDB $\Theta_1$ AUTHOR-Collaborates-in-SIGMOD) $\Theta_3$ (NOT(AUTHOR-Collaborates-in-DASFAA) $\Theta_2$ NOT(AUTHOR-Collaborates-in-DaWaK)) & Community & AND\\
            \hline
        
            \hline     
            \hline
        \rowcolor{yellow!70}\multicolumn{4}{|p{11.5cm}|}{\textbf{IMDb MLN Layers:} ACTOR-Acts-with, ACTOR-Similar-Genre, ACTOR-Similar-AverageRating, DIRECTOR-Similar-Genre, MOVIE-Similar-Rating} \\    
            \hline
            \ref{analysis:IMDbHoMLN1} & ACTOR-Acts-with $\Theta$ ACTOR-Similar-AverageRating & Community  & AND \\
            \hline
            \ref{analysis:IMDbHoMLN2} & NOT(ACTOR-Acts-with) $\Theta_1$ ACTOR-Similar-AverageRating $\Theta_2$ ACTOR-Similar-Genre, Filter = sort-AverageRating & Community & AND \\
            %NOT(ACTOR-Acts-with) $\Theta$ ACTOR-Similar-Genre $\Theta$ ACTOR-Similar-AverageRating & Community & AND (HoMLN) \\
            \hline
             \ref{analysis:IMDbHeMLN1} & ACTOR-Similar-Genre $\Theta$ DIRECTOR-Similar-Genre & Community  & MWM \\
            \hline
            %\ref{analysis:IMDbHeMLN2} & ACTOR-Similar-Genre $\Theta$ MOVIE $\Theta$ DIRECTOR $\Theta$ ACTOR-Similar-Genre & Community & MWM (\underline{HeMLN}) \\
            \ref{analysis:IMDbHeMLN2} & MOVIE-Similar-Rating $\Theta_{1,M}$ ACTOR-Similar-Genre $\Theta_{2,M}$ DIRECTOR-Similar-Genre $\Theta_{3,D}$ ACTOR-Similar-Genre & Community & MWM \\
            \hline
        
            \hline     
            \hline
        \rowcolor{blue!25}\multicolumn{4}{|p{11.5cm}|}{\textbf{COVID HoMLN Layers:} USCOUNTY-Similar-Cases-Period-t1, USCOUNTY-Similar-Cases-Period-t2} \\    
            \hline
            \ref{analysis:Covid1} & Apply $\Psi$ on each layer & Community & N/A \\
        %    \hline
        % \ref{analysis:Covid2} &  Similar-Cases-Period6(Apr16-Apr29), Similar-Cases-Period17(Sep17-Sep30) & Community & none\\
            \hline  
\end{tabular}
%\vspace{-10pt}
\caption{Generated MLN Expression for Each Analysis Objective}
\vspace{-20pt}
\label{table:computation}
%%\end{table}
\end{table}

\section{Knowledge Discovery, Drill-Down, and Visualization}
\label{sec:experiments}

\noindent We computed the results for each analysis objective using the expressions derived and shown in Table~\ref{table:computation} and compare it, where possible, with independently available ground truth. This helps validate both the modeling and analysis aspects of the life cycle approach proposed. We will not focus on the the efficiency of the decoupling approach as it has been established elsewhere~\cite{ICCS/SantraBC17,ICDMW/SantraBC17}. Structure- and semantics-preserving aspects of the decoupling approach allows us to drill-down and show details of experimental results.
%%Versatility should be clear by now from the variety of analysis objectives and re-use of computed results.
%%In-depth analysis after drilling down into the results for \ref{analysis:Air1} to \ref{analysis:IMdbHe-madm} are presented in Section \ref{sec:exp-results} followed by the performance analysis in Section \ref{sec:exp-efficiency}.

%%%%%%%%%%

\subsection{{US-Airline  Analysis Results}}
\label{sec:exp-results}
%We now discuss the analysis results for the US airlines.

\textbf{MLN Details:} Based on the direct flights that were active in February 2018, US-Airline MLN layers are generated whose statistics are shown in Table \ref{tab:USAirlineHoMLNStats}.

\begin{table}[h]
\renewcommand{\arraystretch}{1}
%\vspace{-20pt}
\centering
    \begin{tabular}{|c|c|c|}
        \hline
        \textbf{Layer} & \textbf{Number of Nodes} & \textbf{Number of Edges} \\
        \hline
        American-Direct-Flight & 270 & 746 \\
        \hline
        Southwest-Direct-Flight & 270 & 717 \\
        \hline
        Delta-Direct-Flight & 270 & 688 \\
        \hline
        Frontier-Direct-Flight & 270 & 346 \\
        \hline
        Spirit-Direct-Flight & 270 & 189 \\
        \hline
        Allegiant-Direct-Flight & 270 & 379 \\
        \hline
        \end{tabular}
%\vspace{-10pt}
\caption{US-Airlines HoMLN Layer Statistics}
\label{tab:USAirlineHoMLNStats}
%\vspace{-10pt}
\end{table}

%%\subsection{Drill-Down Analysis of Results}

%The HoMLN described in Section \ref{sec:analysis} was built for 290 US cities using \textit{flights active until February 2018}. The layer(airline)-wise statistics have been shown in Table \ref{table:USAirlineHoMLNStats}.

\noindent \ref{analysis:Airline1}
\textbf{For American, Southwest, Spirit, Delta, Frontier and Allegiant Airlines rank the top five cities, that  provide  the maximum  coverage}.
%\ref{analysis:Airline1} \textbf{Rank the top five cities for each airline, that have the maximum coverage.} 
Based on the expression derived, we computed the closeness centrality for each layer. We ranked the cities in each layer according to their closeness centrality value. Top 5 hubs (\textit{higher rank, fewer flights required for coverage, more central city})) were identified for each airline.  {\em For all 6 airlines, the ground truth obtained from~\cite{hubs} matched our results.} In Table  \ref{table:A1} we have listed top 5 hubs for 4 airlines. As a drill-down byproduct, it is interesting to see common hubs (highlighted) between airlines which is also verified by the ground truth.

\begin{table}[h]
\centering
\vspace{-15pt}
\subfloat[]{

    \begin{tabular}{|c|}
        \hline
        \textbf{American} \\
        \hline
        \hline
        \textit{Dallas} \\
        \hline
        \textit{Chicago} \\
        \hline
        Charlotte \\
        \hline
        Philadelphia \\
        \hline
        Phoenix \\
        \hline
        \end{tabular}
        \label{table:A1-results-AA}
}
\subfloat[]{

    \begin{tabular}{|c|}
        \hline
        \textbf{Southwest} \\
        \hline
        \hline
        \textit{Chicago} \\
        \hline
        \textbf{Denver} \\
        \hline
        Baltimore \\
        \hline
        \textit{Dallas}  \\
        \hline
        \texttt{Las Vegas}  \\
        \hline
        \end{tabular}
        \label{table:A1-results-SW}
}
\begin{comment}

\subfloat[]{

    \begin{tabular}{|c|}
        \hline
        \textbf{Delta} \\
        \hline
        \hline
        Atlanta \\
        \hline
        Minneapolis  \\
        \hline
        Detroit \\
        \hline
        Salt Lake City \\
        \hline
        New York \\
        \hline
        \end{tabular}
        \label{table:A1-results-DL}
}
\end{comment}
\subfloat[]{

    \begin{tabular}{|c|}
        \hline
        \textbf{Frontier} \\
        \hline
        \hline
        \textbf{Denver} \\
        \hline
         \texttt{Orlando} \\
        \hline
         Austin \\
        \hline
         \texttt{\textbf{Las Vegas}} \\
        \hline
         Philadelphia \\
        \hline
        \end{tabular}
        \label{table:A1-results-FT}
}
\subfloat[]{

    \begin{tabular}{|c|}
        \hline
        \textbf{Spirit} \\
        \hline
        \hline
        Ft. Lauderdale \\
        \hline
         \texttt{\textbf{Las Vegas}} \\
        \hline
         \texttt{Orlando} \\
        \hline
         Detroit \\
        \hline
         \textit{Chicago} \\
        \hline
        \end{tabular}
        \label{table:A1-results-SP}
}
\begin{comment}
\subfloat[]{

    \begin{tabular}{|c|}
        \hline
        \textbf{Allegiant} \\
        \hline
        \hline
        Orlando \\
        \hline
        Tampa  \\
        \hline
         Las Vegas \\
        \hline
         Phoenix \\
        \hline
         Fort Myers \\
        \hline
        \end{tabular}
        \label{table:A1-results-AG}
}
\end{comment}
%\qquad

\vspace{-5pt}
\caption{\ref{analysis:Airline1}: Cities With Maximum US Travel Coverage}
\label{table:A1}
%\vspace{-10pt}
    
\end{table}

\noindent 
\ref{analysis:Airline3} \textbf{Predict which city (taking its population into consideration) could be selected as the next hub(s) for Allegiant Airlines to expand  its coverage and avoid competition with other airlines.} 
%\ref{analysis:Airline3} {\bf Predict next hub for an airline.}

%Here, Allegiant is the airline which is considering expansion based on the objective. All remaining airlines are chosen as competitors.

%to determine the next promising city for Spirit to expand its operations, we considered two different competitor airlines - Southwest (major), Allegiant (minor), shown in Table \ref{table:A3-results-SP}.

\begin{wraptable}{l}{0.35\linewidth}
\small
\centering
\vspace{-10pt}
    \begin{tabular}{|c|c|}
        \hline
        City & Pop. (2019)\\
        \hline
        \hline
        \textbf{Grand Rapids} & 198,401 \\
        \hline
        Elko & 20,304\\
        \hline
        Montrose & 19,238\\
        \hline
        \end{tabular}
        \label{table:A3-results-AGvALL}
%}
\caption{\ref{analysis:Airline3}: Next Allegiant hub by Rank} %Expansion Cities}
\label{table:A3}
\vspace{-20pt}
\end{wraptable}

%The expression generated can be interpreted as follows. Identify cities for Allegiant that are a) not yet a hub, b) have high coverage, i.e. high values of closeness centrality which helps to reduce cost of expansion and c) do not have large operations, i.e. low closeness centrality for the competitor airlines which helps to minimize competition.

%This translates to the computation: DegreeHubs(Target) - (ActualHubs(Target) $\cup$ ( \textit{DegreeHubs(Target) AND DegreeHubs(Competitor) )}.
%From the high closeness centrality cities of the target airline, we removed the actual hubs first, followed by all those cities that are also high closeness in each of the competitor airlines. We then ranked this set of cities based on their \textit{population}.

On computing the expression, from the high closeness centrality cities of the Allegiant airline, we eliminated all those cities that are also high closeness in each of the competitor airlines. From this set of cities, we ranked those cities that are not currently Allegiant hubs, based on their \textit{population}. This information is available from the \textit{City} relation (Figure \ref{fig:USAirline-Model} (b)) that was obtained as a by-product of the EER $\rightarrow$ MLN process.
%The pruned set was further filtered by using additional external information like population to rank them. %%bring in the aspect of ticket sales likelihood.
Table \ref{table:A3} shows the resulting set of cities where Allegiant Airline can potentially expand its operations. %, based on closeness centrality, when the remaining 5 airlines are considered as competitors, \textit{separately}. 
We validated our result by the fact that \textit{\textbf{Grand Rapids has been converted to a hub by Allegiant as of July 6, 2019}} \cite{AG-hub}. %This shows that \textit{MLN analysis results can be effectively used (augmented with additional information, where needed)} to make real-world business decisions. 

\subsection{DBLP Analysis Results}

%\noindent\textbf{A8: Find  the  co-author groups who have  published in high ranked conferences. }

\textbf{MLN Details:} Table \ref{table:DBLPMLNStats} provides the statistics of the different DBLP MLN layers. Publications from 2001 to 2018 were considered for generating the first three layers - AUTHOR-Collaborates-With, PAPER-Same-Conference and YEAR-Same-Interval. For the generation of remaining 4 layers, the respective conference publications were selected from 2003-2007. Louvain  algorithm was used to generate the layer-wise communities \cite{DBLP:Louvain}.

\begin{table}[h!t]
\renewcommand{\arraystretch}{1}
\centering
%\vspace{-10pt}
    \begin{tabular}{|c|c|c|c|}
        \hline
        
        \textbf{Layer} & \textbf{Number of Nodes} & \textbf{Number of Edges} \\
        \hline

        AUTHOR-Collaborates-With & 16,918 & 2,483 \\
        \hline

        PAPER-Same-Conference & 10,326 & 12,044,080 \\
        \hline

        YEAR-Same-Interval & 18 & 18 \\
        \hline

        AUTHOR-Collaborates-in-VLDB & 5116 & 3912 \\
        \hline

        AUTHOR-Collaborates-in-SIGMOD & 5116 & 3303 \\
        \hline

        AUTHOR-Collaborates-in-DASFAA & 5116 & 1519 \\
        \hline

        AUTHOR-Collaborates-in-DaWaK & 5116 & 679 \\
        \hline

        \end{tabular}
\caption{DBLP MLN Statistics}
\label{table:DBLPMLNStats}
%\vspace{-15pt}
    
\end{table}

%\subsection{DBLP HeMLN Analysis Results}

\begin{figure}[h]
   \centering
%   \vspace{-10pt}
   \includegraphics[width=0.8\linewidth]{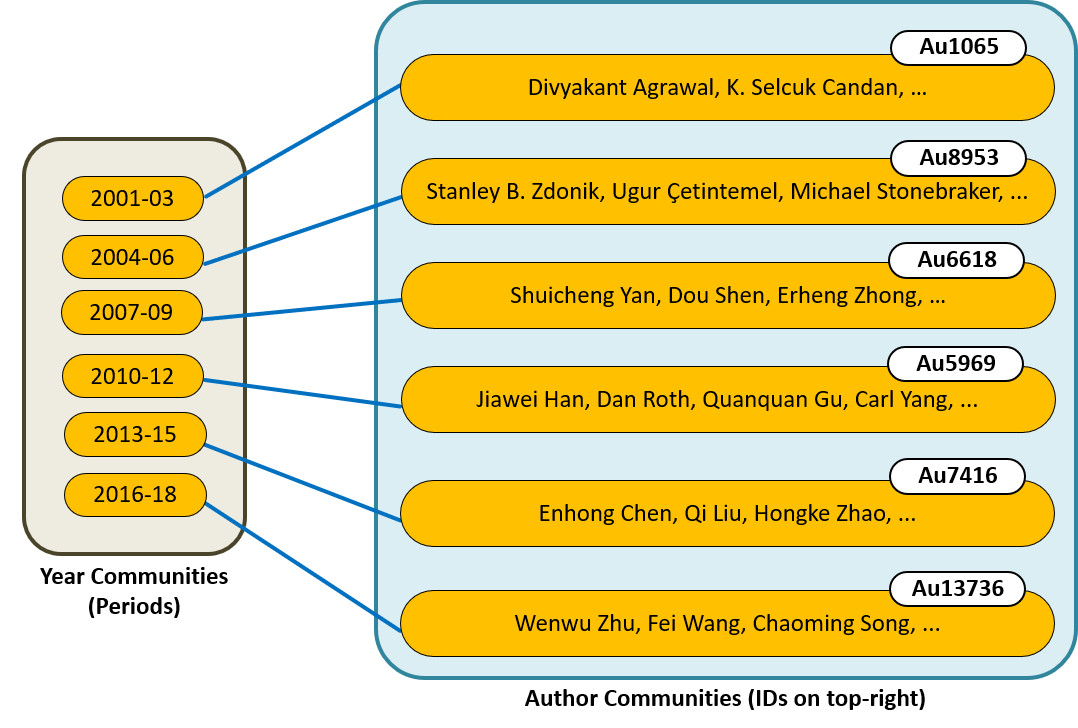}
   \vspace{-5pt}
   \caption{\ref{analysis:DBLPHeMLN1}: Most Active Co-Author Groups}
   \label{fig:DBLP-YAu-we-MWM}
%   \vspace{-10pt}
\end{figure}

\noindent 
\ref{analysis:DBLPHeMLN1} \textbf{For each 3-year interval group, find the most actively publishing strong author collaboration groups}.
%\textbf{\ref{analysis:DBLPHeMLN1}: For each disjoint 3-year period, find the strong co-author groups who were most actively publishing.}
%For each of 6 conferences, we obtained most cohesive author (Au) communities that also has high interactions among themselves. In Figure \ref{fig:DBLPHe-PAu-wd}, paper communities such as ICDM and DAWAK has paired to more than one author communities showing each author community is equally cohesive with these conferences. It can be seen that, distinguished author like George Karypis is associated with ICDM.(some fact about George Karypis) Unlike Maximum weighted bipartite matching, WBC algorithm can give out more than one matching if both matchings have equal weights. This can be seen as authors like Rajeev Rastogi, Minos N. Garofalakis in author community(AU\_10513) are associated with both SIGMOD and VLDB conferences. Similarly 3 author communities(AU\_2241, AU\_9906, AU\_12263) are associated with DAWAK.
As per the expression, on applying MWM on the community bipartite graph created with all Paper and Author communities, we obtained 6 community pairings for the \textit{co-author groups who have published most number of times in each 3-year period}, (shown in Figure \ref{fig:DBLP-YAu-we-MWM} with list of few prominent authors.) This visualization was accomplished by the drill-down of raw results with the help of relations obtained earlier in Figure \ref{fig:DBLP-MLN} (b). The author community ids have been shown that are generated by the community detection algorithm in the $\Psi$ phase of the decoupling approach.  Quality of these results are validated by the following facts,

%\textbf{To be revised}

%Different co-author groups were active across each 3-year period. For instance,

\begin{itemize}
    \item For the period from 2004 to 2006, the most active group included prominent researchers like \textbf{Stanley B. Zdonik (ACM Fellow from 2006), Ugur Çetintemel and Michael Stonebraker (Turing Award Recepient, 2014} who co-authored papers on stream processing engines in this period garnering over \textbf{2000 citations}. 
    \item From 2010 to 2012, \textbf{Jiawei Han}, one of the most cited computer science researchers (194752 citations), was involved in more than \textbf{200 publications}. In the beginning of this period, he had even won the McDowell Award, the highest technical award made by IEEE.
\end{itemize}

Such insightful results can be further drilled-down to find active periods of co-author subgroups, research labs and universities.

\vspace{5pt}

\begin{figure}[h]
   \centering
%   \vspace{-10pt}
   \includegraphics[width=\linewidth]{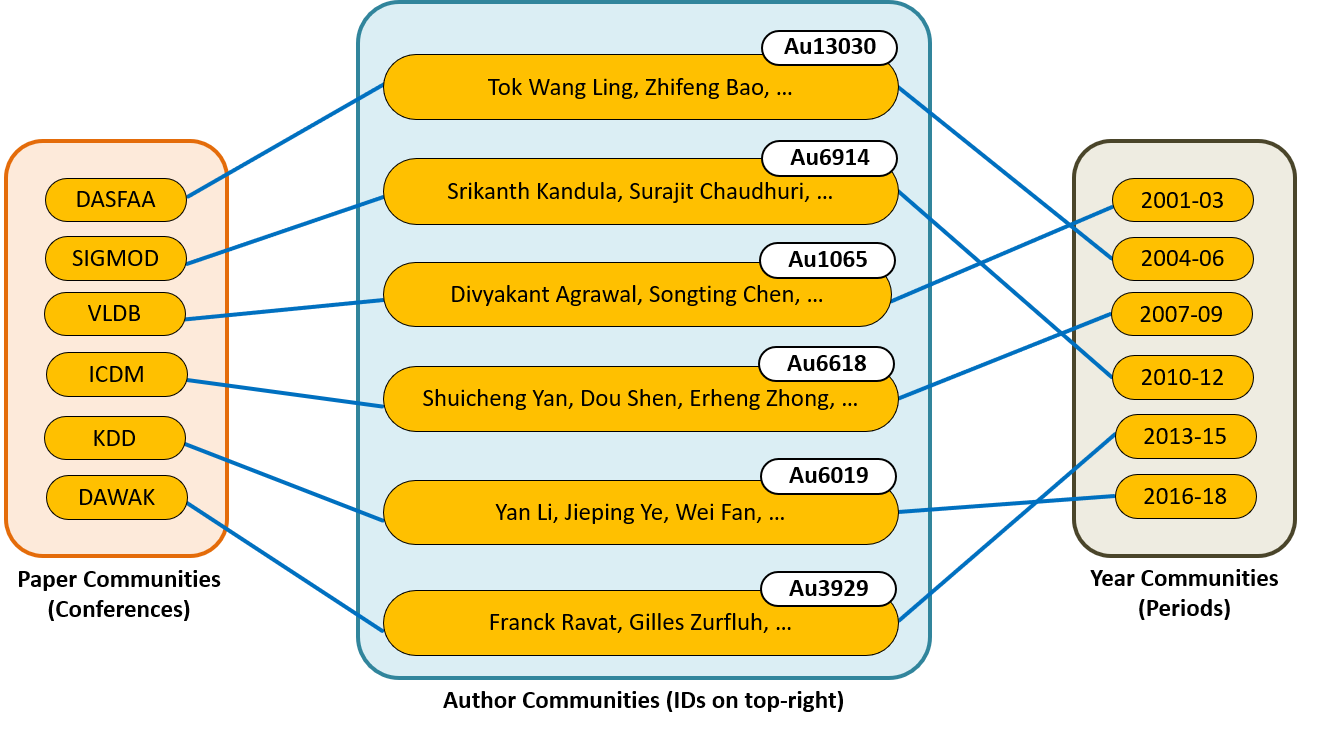}
   \vspace{-15pt}
   \caption{\ref{analysis:DBLPHeMLN2}: Active Periods for Popular Conference-wise Co-Author Groups}
   \label{fig:DBLP-PAuY-we-MWM}
%   \vspace{-10pt}
\end{figure}

\noindent \textbf{\ref{analysis:DBLPHeMLN2} For each conference-based paper group, find the most popular author collaboration group and further for each of them identify their most active 3-year interval group(s)}.
%For the most popular collaborators in each conference, identify the 3-year period(s) when they were most active. } 
%In this analysis, \textit{most popular author communities for each conference was established and for each popular author community we obtained the year in which they were most active in}. In the figure \ref{fig:DBLP-PAuY-we-MWM}, each paper community (i.e. conference) has one most popular group of co-authors associated. 
In order to generate the required communities, based on the expression in Table \ref{table:computation}, the \textit{most popular author groups} for \textit{each conference} are obtained by MWM (first composition). The matched 6 author communities are carried forward to find the \textit{disjoint year periods} in which they were \textit{most active} (second composition.) 6 communities are obtained (path shown by \textbf{\textcolor{blue}{bold blue lines}} in Figure \ref{fig:DBLP-PAuY-we-MWM}.) 

Few prominent names have been drilled-down and shown in Figure \ref{fig:DBLP-PAuY-we-MWM} based on citation count (from Google Scholar profiles.) For example, for \textit{SIGMOD, VLDB and ICDM} the most popular researchers include \textbf{Srikanth Kandula (15188 citations), Divyakant Agrawal (23727 citations) and Shuicheng Yan (52294 citations)}, respectively who were active in different periods in the past 18 years.

\begin{figure}[ht]
    \centering
    %\vspace{-10pt}
  \includegraphics[width=0.8\linewidth]{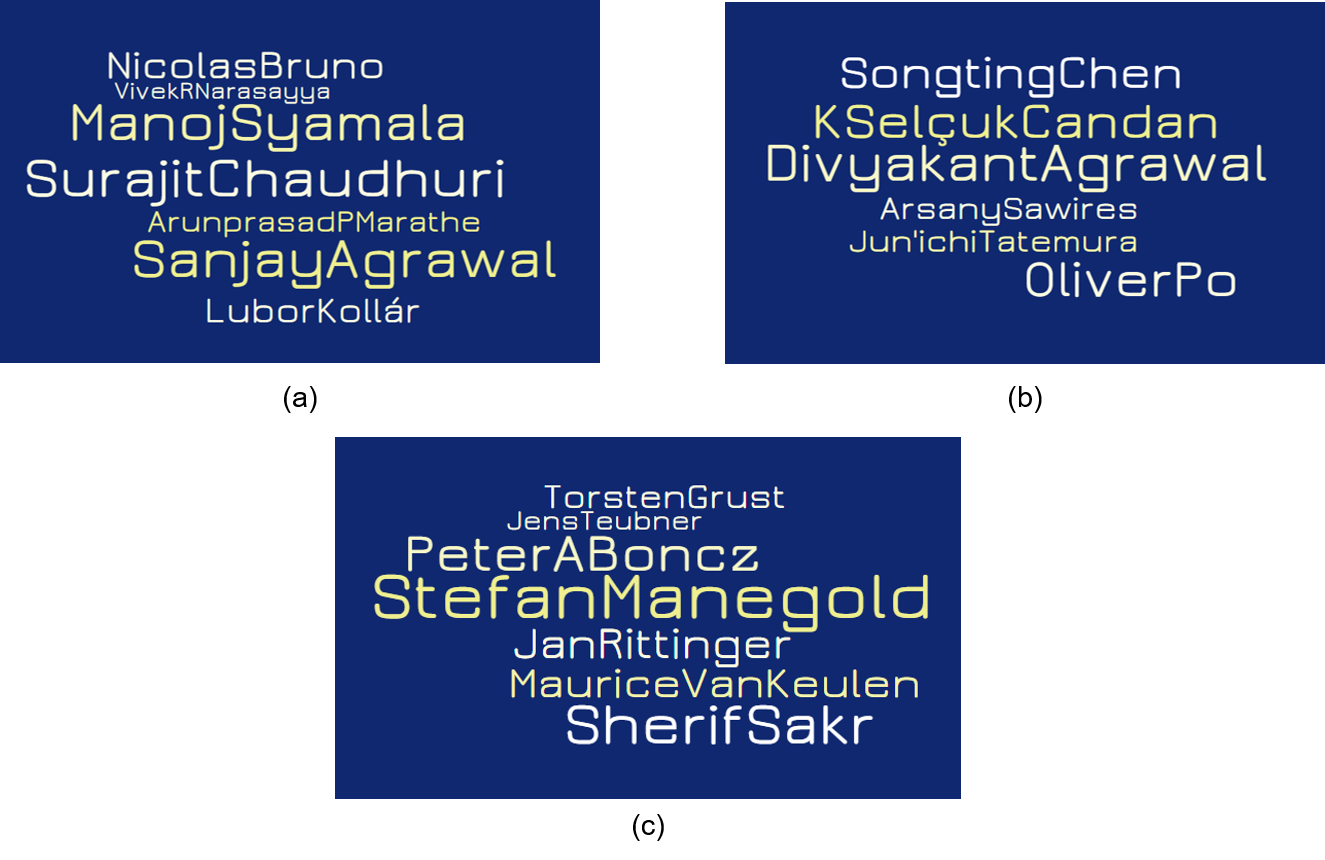}
  \vspace{-15pt}
    \caption{\ref{analysis:DBLPHoMLN1}: Author Groups with High Quality Research}
    \label{fig:Drill-down}
    \vspace{-15pt}
\end{figure}

\noindent \textbf{\ref{analysis:DBLPHoMLN1} Identify author collaboration groups who have published in conferences VLDB and SIGMOD, but have never published in conferences DASFAA and DaWaK\footnote{Note that these conferences can be changed and analyzed using the \textit{same} expression, but with different layers. Parameterization of analysis objectives is one of the advantages of this approach!}.} 102 communities are obtained after computing the expression. Drill-down results have been shown in Figure \ref{fig:Drill-down} for few well-known groups most of whose members had collaborated on a paper that was published in both VLDB and SIGMOD (high ranked), but never in DASFAA or DaWaK (low to medium ranked). %in the period from 2003 to 2007.

There is a high probability that the \textbf{work done by these groups is not only of good quality but also widely accepted}. This claim is validated through the following facts:

\begin{itemize}
    \item Figure \ref{fig:Drill-down} (a) community has researchers like \textbf{Surajit Chaudhari} who won the \textbf{VLDB 10-Year Best Paper Award (2007)} with \textbf{Vivek Narasayya} and \textbf{VLDB Best Paper Award (2008)} with \textbf{Nicolas Bruno}, apart from winning \textbf{ACM SIGMOD Contributions Award (2004)}.
    \item Figure \ref{fig:Drill-down} (b) has researchers like \textbf{Divyakant Agrawal} who has \textbf{24000+ citations} (Google scholar).
    \item \textbf{Peter A. Boncz and Stefan Manegold} from Figure \ref{fig:Drill-down} (c) group not only published a \textbf{highly cited paper} (350+ citations for MonetDB/XQuery) in SIGMOD 2006, but also have won the \textbf{VLDB 10-year award}.
\end{itemize}

%%%%%%%%%%%%%%%%%%%%

\subsection{{IMDb Analysis Results}}

\textbf{MLN Details:} For IMDb MLN,  we extracted, for the top 500 actors, the movies they have worked in (7500+ movies with 4500+ directors). The actor set was repopulated with the co-actors from these movies, giving a total of 9000+ actors. As explained for DBLP, the relationship attribute parameters in the EER model help in quantifying the similarity of actors and directors based on movie genres they have worked in. A vector was generated with the number of movies for each genre he/she has acted-in/directed. In order to consider the similarity with respect to \textit{frequency of genres}, two actors/directors are connected if the Pearsons' Correlation between their corresponding genre vectors is at least 0.9 (Other values can also be used based on similarity strength.) Widely used Louvain method~\cite{DBLP:Louvain} is used to detect layer-wise communities ($\Psi$). Table \ref{table:IMDbMLNStats} provides the layer statistics of the generated MLN.

\begin{table}[h!t]
\renewcommand{\arraystretch}{1}
\centering
%\vspace{-10pt}
    \begin{tabular}{|c|c|c|}
        \hline
      \textbf{Layer}  & \textbf{Number of Nodes} & \textbf{Number of Edges} \\
        \hline
        ACTOR-Acts-with & 9485 & 45,581\\
        \hline 
        ACTOR-SimilarAverageRating & 9485 & 13,945,912\\
        \hline
        ACTOR-Similar-Genre & 9485 & 996,527\\
        \hline
        DIRECTOR-Similar-Genre & 4510 & 250,845\\
        \hline
        MOVIE-Similar-Rating & 7951 & 8,777,618 \\
        \hline
        \end{tabular}
\caption{IMDb MLN Statistics}
\label{table:IMDbMLNStats}
%\vspace{-10pt}
    
\end{table}

\noindent {\bf \ref{analysis:IMDbHoMLN1} 
%Cluster groups of co-actors who have acted in similarly rated movies.}
Cluster actors who have acted together and have a similar average rating}.
2430 actor groups with similar average ratings were detected in which \textit{most of the actor pairs} have worked with each other. Few drill-down observations on the results:
\begin{itemize}
    \item For the most popular average actor rating, [6-7), the largest co-actor groups were from Hollywood (876 actors), Indian (44 actors), Hong Kong (12 actors) and Spanish (9 actors) movies.
    \item Among the Hollywood movie based groups, the top group included co-actors  {Al Pacino, Robert De Niro, and Will Smith}. Pacino and  De Niro acted together in famous movies like Heat and Godfather Part II. Interestingly, co-actors from less known movies, such as Smith and De Niro, in Shark Tale also come up.
    \item Famous Bollywood stars like {Amitabh Bachchan}, {Shah Rukh Khan} belonged to largest top rated Indian group. They acted together in many highly rated bollywood movies. 
    \item {Jackie Chan} (along with other lesser known actors) was among the prominent actors from the co-actor group from Hong Kong.
\end{itemize}
%\end{comment}

%\forceindent In case of \ref{analysis:IMdbHo-2}, 592 actor groups who opt for \textit{similar movie genres} are detected across \textit{different average rating ranges}. As a part of this analysis, the group corresponding to Action, Adventure and Sci-Fi genres and an average rating of [6-7) had actors like Robert Downey Jr., Mark Ruffalo and Scarlett Johansson. This is primarily because these actors/actresses have been part of the Marvel Cinematic Universe movies over the past 10 years.

%\begin{table}{r}{0.5\linewidth}
%\end{wraptable}

\noindent \textbf{\ref{analysis:IMDbHoMLN2}
%Find highly-rated actors who have worked in similar genres, but have not acted together.}
Find the groups of actors who have never acted together, but are highly rated on an average and have worked in similar genres}. 
Following the expression, we detected 900 groups of actors most of whom have not worked together but have similar genre preferences and average rating. From the results, we drilled down into the communities that corresponded to \textit{high average rating} and have listed a few \textit{recognizable actors} along with \textit{prominent genres} from those communities in Table \ref{table:actorcollab}. Out of these, as per reports in 2017,  there had been {\bf talks of casting Johnny Depp and Tom Cruise in pivotal roles in Universal Studios' cinematic universe titled Dark Universe} \cite{dark-universe}.

\begin{table}[h!t]
\vspace{-5pt}
\renewcommand{\arraystretch}{1}
\centering
    \begin{tabular}{|p{4cm}|p{5cm}|}
        \hline
        \textbf{Actors} & \textbf{Common Prominent Genres} \\
        \hline
        Dafoe, Crowe & Action, Crime\\%, Drama \\
        \hline
        Swank, Winslet & Drama \\
        \hline
        Hanks, Witherspoon, Diaz & Comedy, Romance\\%, Drama\\
        \hline
%        Anne Hathaway, Salma Hayek & \\
%        \hline
        \texttt{\textbf{Depp, Cruise}} & \texttt{Adventure, Action}\\%, Drama \\
        \hline
%        Brad Pitt, Will Smith & \\
%        \hline
        DiCaprio, Gosling & Crime, Romance\\%, Drama\\
        \hline
        Cage, Banderas & Action, Thriller \\%, Drama
        \hline
        Grant, Hudson, Stone & Comedy, Romance \\%Drama

        \hline
        \end{tabular}
%\vspace{-5pt}        
\caption{\ref{analysis:IMDbHoMLN2}:  Highly rated genre actors who have \textbf{not co-acted}.}
\label{table:actorcollab}
%\vspace{-10pt}
\end{table}

\noindent {\bf \ref{analysis:IMDbHeMLN1} 
%Find groups of actors and directors who collaborate together.} 
Identify genre-based groups of  actors  and  directors  having  strong collaborations}. 
On computing the expression, 49 similar genre-based community pairs are obtained, where {most actor-director pairs have interacted with each other at least once}. Intuitively, a group of actors that prominently works in some genre (say, Drama, Action, Romance, ...) must pair up with the group of directors who primarily make movies in the \textit{same genre}.  

 \begin{figure}[h]
   \centering
%   \vspace{-10pt}
   \includegraphics[width=0.6\linewidth]{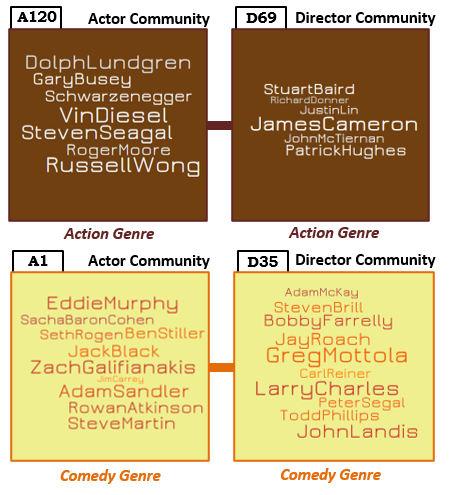}
   \vspace{-5pt}
   \caption{\ref{analysis:IMDbHeMLN1}: Genre-based Actor-Director Groups}
   \label{fig:IMDb-HeMLN-AD}
   \vspace{-22pt}
\end{figure}

In Figure \ref{fig:IMDb-HeMLN-AD} we have drilled-down and visualized the community pairings for the Action and Comedy genres with few famous actors and directors from each community. Such pairings may help production houses to sign up actors and directors for different movie genres. Recently, \textbf{Vin Diesel signed up for Avatar 2 and 3 (Action movie) which is being directed by James Cameroon and this will be the first time they will be collaborating}~\cite{avatar2}. Interestingly, even though they did not work together ever, we paired them together in the groups that corresponded to the Action genre on the basis of \textit{high interaction among other similar actors and directors}. %~\footnote{This pairing is not shown in the Fig.\ref{fig:imdb-hemln} due to space constraints}. %Thus, potential actor-director collaborations can be explored using MLN analysis.

%\end{comment}

%\forceindent For \ref{analysis:IMdbHe-mad}, a different acyclic ordering for same 3 layers is computed, generating 5 M-A-D community matches. There were no M-A matches in this case, that did not get extended.

%%{\bf A7: Find groups of actors and directors who collaborate together in highly rated movies.} We obtained the most popular actor (A) and director (D) community for each movie rating (M) community, that also have high interaction among them, through 3 iterations of MWBM. In Figure \ref{fig:imdb-hemln} (b) the \textit{most popular actor and director groups for [6-7) movie rating are from different genres}. Even though few actor-director pairs from these two have collaborated on a few movies, it can be seen from Figure \ref{fig:imdb-hemln} (a) that D91 pairs (has maximum interaction) with A94. Thus, validating the absence of pairing between D91 and A144. However, in case of Figure \ref{fig:imdb-hemln} (c), \textit{both the popular groups for [7-8) rating are from Drama genre and many actor-director pairs have collaborated on many movies} like \textit{Leonardo DiCaprio, Kate Winslet with Sam Mendes for Revolutionary Road}, \textbf{Sean Penn with Gus Van Sant for Milk} and so on. Thus, the popular groups A175 and D106 are also paired with each other.

\noindent \textbf{\ref{analysis:IMDbHeMLN2}:
%Identify, for each movie rating, the most popular genre-based actor and director  groups who have strong collaborations
Identify, for each movie rating group the genre-based most popular actor and most popular director groups. From this result, find the actor and director groups having strong collaborations.} %We obtained the HeMLN community between most popular actors (A) and highly rated movies (M) using the expression in Table~\ref{table:computation}. 
%OLD VERSION
%When finding the communities across three layers, using the expression in Table~\ref{table:computation}, we first tried to combine results each of two layers with that of a common layer. Figure~\ref{fig:imdb-hemln} (b) shows the  results of one such combination, where actors (community A144) and directors (community D91) is paired with movies (community M3). However, we see that most popular actor and director groups for [6-7) movie rating (represented by M3) do not have many interactions. Even though few actor-director pairs from these two have collaborated on a few movies, it can be seen from Figure \ref{fig:imdb-hemln} (a) that D91 (community id in layer D) pairs (has maximum interaction) with A94, thus validating the absence of pairing between D91 and A144 in this result.
% NEW VERSION
\begin{figure}[h]
%\vspace{-20pt}
   \centering \includegraphics[width=0.65\columnwidth]{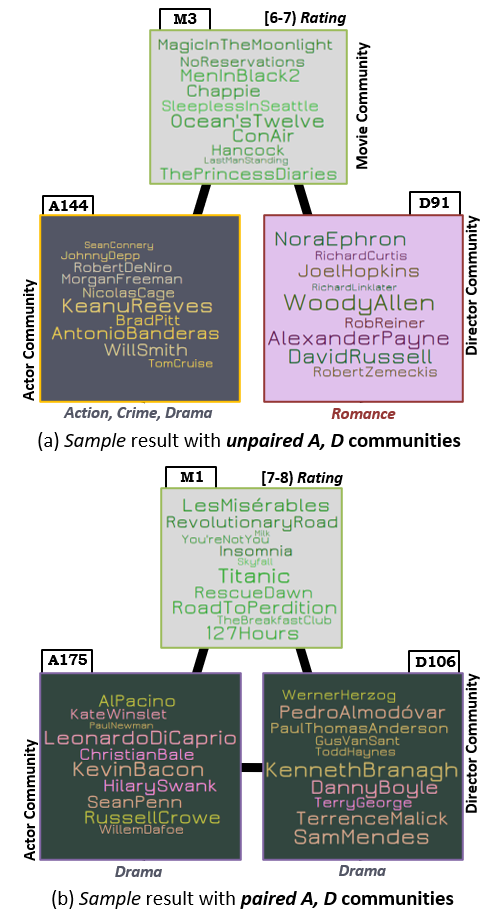}
  \vspace{-7pt}
   \caption{\ref{analysis:IMDbHeMLN2}: \textit{Sample} HeMLN community match results.}
   \label{fig:imdb-hemln-ADM}
\vspace{-18pt}
\end{figure}
When finding the communities across three layers, using the expression in Table~\ref{table:computation}, we first combine results of each of two layers (ACTOR-Similar-Genre, DIRECTOR-Similar-Genre) with that of common layer (MOVIE-Similar-Rating) to find most popular genre-based group for each movie rating. %, that is ACTOR-Similar-Genre \textbf{--} MOVIE \textbf{--} DIRECTOR.
Figure~\ref{fig:imdb-hemln-ADM} (a) shows  drill-down results of one such \textit{intermediate} combination, where actors (community A144) and directors (community D91) are paired with movies (community M3). However, the most popular actor and director groups for [6-7) movie rating (represented by M3) do not have many interactions among them as they \textit{belong to different dominant genre groups}. %Even though few actor-director pairs from these two have collaborated on a few movies, it can be seen from Figure \ref{fig:imdb-hemln} (a) that D91 (community id in layer D) pairs (has maximum interaction) with A94, thus validating the absence of pairing between D91 and A144 in this result.

%This result motivated us to couple the results of the three layers. Here the interactions between actors and movie communities are calculated, then interactions between movie and director communities are calculated and finally interactions between directors and actors are calculated. 

Finally, the interactions between DIRECTOR-Similar-Genre communities and ACTOR-Similar-Genre communities are calculated to complete the analysis expression listed in Table \ref{table:computation}. 
%Only the communities for which the coupled paths (\textbf{ACTOR-Similar-Genre \textbf{--} MOVIE \textbf{--} DIRECTOR \textbf{--} ACTOR-Similar-Genre}) are obtained are retained as results.
%Then, we composed a 3-layer HeMLN using the results of the previous step for the Movie layer with collaborative directors (D). Figure~\ref{fig:imdb-hemln} (b) shows one of the results: that \textit{most popular actor and director groups for [6-7) movie rating are from different genres}.  This also illustrates the community refinement as more layers are included. 
%The result of the above computation was further composed with the A layer results using the D layer results to see whether a 2-way strong coupling between A-M-D can be extended to a 3-way strong coupling between A-M-D-A. 
Only one HeMLN community drilled-down and visualized in Figure~\ref{fig:imdb-hemln-ADM} (b) was obtained. %This shows the capability of the HeMLN community detection to identify communities that \textbf{strongly interact cyclically in all 3 layers.}
The drill-down of Figure \ref{fig:imdb-hemln-ADM} (b) indicates, \textit{both the popular groups for [7-8) movie rating are from Drama genre and many of these actor-director pairs have collaborated on many movies}, such as {\bf Leonardo DiCaprio, Kate Winslet with Sam Mendes for Revolutionary Road, Sean Penn with Gus Van Sant for Milk} and so on. Thus, popular groups A175 and D106 paired up with each other.

Most importantly, it is possible by drilling-down into the results to flesh out \textbf{potential actor-actor or actor-director collaborations} based on \textit{identifying the missing links for high degree nodes in the generated HeMLN communities}. One such combination is \textbf{DiCaprio-Swank-Mendes} who \textit{never worked together even though most of their movies belong to highly rated drama genre}.

%\textcolor{red}{Although it is present in the text, a clearer definition of what is uncoupled and what  is coupled. Also the terms might be confused with the "decoupling approach".}

%Out of the 9 A-M-D community matches, only 2 matches are such where the actor and director communities also have high interaction among them, thus producing 2 A-M-D-A community matches. Figure \ref{fig:amda} (b) shows one of those two matches, whereas Figure \ref{fig:amda} (a) shows one of the remaining 7 A-D-M matches where the A and D communities \textit{did not get coupled}. %figure \ref{fig:madm} shows out of the 5 M-A-D acyclic matches, only 3 got updated (through consistent matches between D and M layer in final iteration) to produce 3 cyclic M-A-D-M matches. The remaining two were retained as partial elements due to inconsistency in the match for the last iteration. 

\subsection{Covid-19 Analysis Results}
\label{sec:CovidAnalysis}

%\sharma{4/14/21}{is it possible to include edge numbers as well?}

\textbf{MLN Details:} Each node in a layer corresponds to a county (3141 nodes.) County nodes are connected as a clique if they have the similar increase, using several bands from spike ($>$ 100\% increase) to big dip (100\% decrease) and a few in between, in the number of Covid new cases/deaths/hospitalizations/... and hence varies.

Based on the analysis objective, 2 disjoint intervals (each ranging from 1 to 30 days) are selected -- either arbitrarily or before and after based on an event (e.g., July $4^{th}$, Thanksgiving) to visually understand the  effect of Covid between the chosen two intervals. This is translated to the generation of layers for each interval, respectively, where the 3141 US counties with similar number of new cases are connected from each interval. As per the expression, communities are detected (using Louvain \cite{DBLP:Louvain}) for the individual layers to find the geographical regions (or counties) that have the \textit{same} percentage of increase or decrease (using bands) and 2 maps are displayed side-by-side using different colors and counties within the same band having the same color. The colors range from \textbf{\color{red}spike} to a \textbf{\color{green}big dip} in the number of new cases/deaths/hospitalizations/.... Different interval selections can be made based around the event of interest to analyze and visualize the effects through the live dashboard that makes use of the multilayer network architecture underneath\footnote{Dashboard: \url{https://itlab.uta.edu/cowiz/}, Youtube Video: \url{https://www.youtube.com/watch?v=4vJ56FYBSCg}}. The effects of two major events visualized in this paper are discussed below. More details can be found in \cite{ICDE2021Demo/Cowiz}.

 \begin{figure}[h]
  %% \vspace{-20pt}
   \centering
   \includegraphics[width=\linewidth]{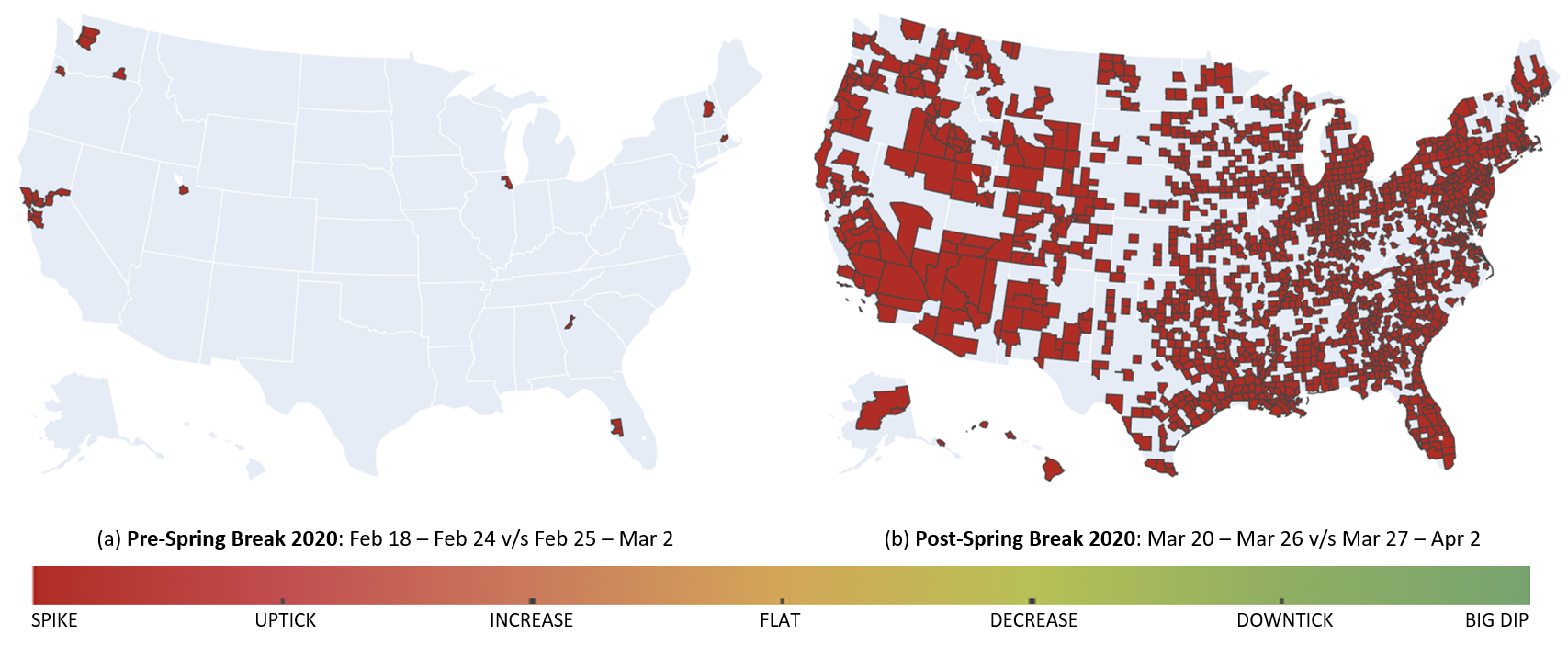}
   \caption{\ref{analysis:Covid1} \textbf{(i)} \textbf{\color{red}Spike} in Cases due to the 2020 Spring Break}
   \label{fig:springbreak}
  %% \vspace{-20pt}
\end{figure}

\noindent \ref{analysis:Covid1} \textbf{(i)} \textbf{Visualize how the  geographical regions corresponding to the clusters of US counties with rise in  daily confirmed cases shift in the consecutive 7-day intervals pre and post \textit{2020 Spring Break}.}  
%\noindent \textbf{(i)} \textbf{Comparison of consecutive 7-day periods pre and post the \textit{2020 Spring Break}}:
For most US counties, the spring break was till the third week of March in 2020. For the \textit{pre} spring break layer, the 7-day intervals used were Feb18-Feb24 and Feb25-Mar2. For the \textit{post} spring break layer, the 7-day intervals used were Mar20-Mar26 and Mar27-Apr2. The drilled-down results have been visualized in Figure \ref{fig:springbreak} that show how \textbf{post the spring break there was a spike in the number of daily cases in counties across the US.} Various reports attributed this massive surge due to the widespread travel to popular tourist destinations during the break leading to \textbf{crowds and non-adherence to social distancing norms} \cite{springbreak2020-1,springbreak2020-2,springbreak2020-3}.

 \begin{figure}[h]
   \centering
   \includegraphics[width=\linewidth]{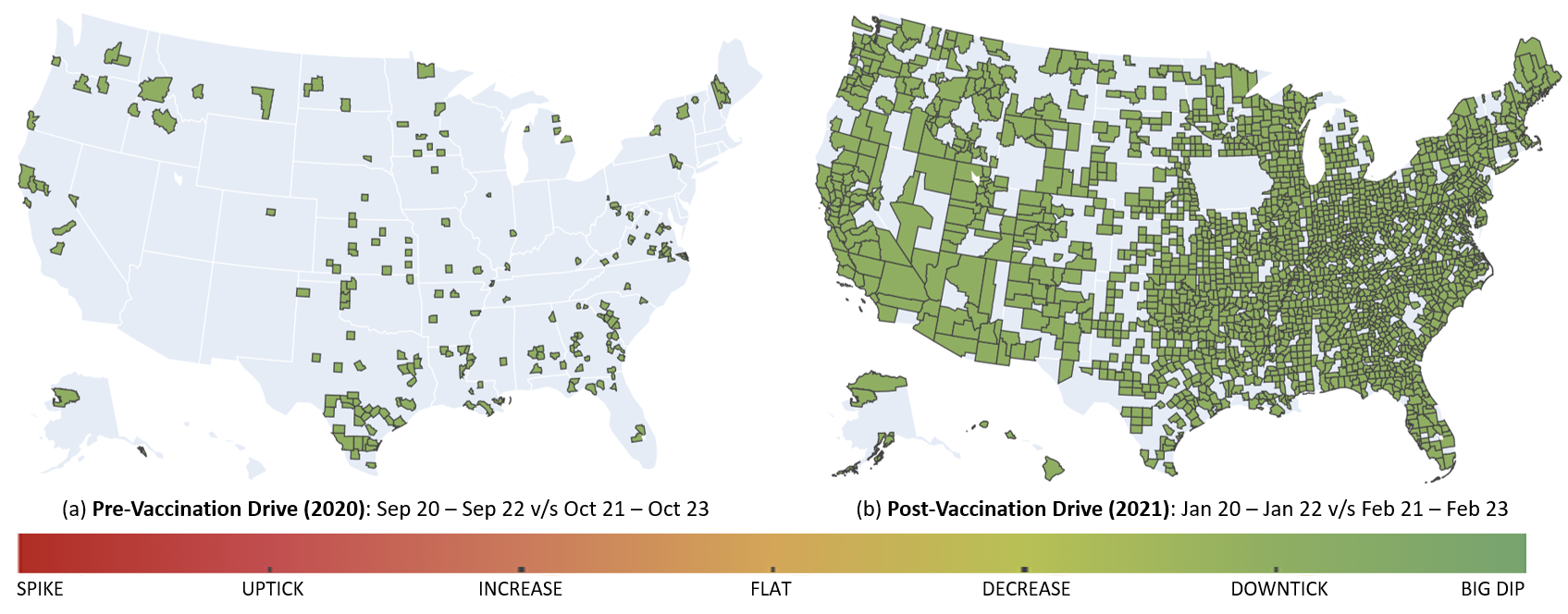}
   \caption{\ref{analysis:Covid1} \textbf{(ii)} Positive Impact of the Vaccination Drive in the US}
   \label{fig:vaccine}
   \vspace{-15pt}
\end{figure}

\noindent \ref{analysis:Covid1} \textbf{(ii)} \textbf{Visualize how the geographical regions corresponding to the clusters of US counties with decline in daily confirmed cases shift in month-apart 3-day periods pre and post the \textit{Vaccination Drive}}: The vaccination drive in the US began from December 14, 2020 \cite{vaccine-0}. For the \textit{pre} vaccination drive layer, the 3-day intervals considered were Sep20-Sep22 and Oct21-Oct23 in 2020. For the \textit{post} vaccination drive layer, the 3-day intervals were Jan20-Jan22 and Feb21-Feb23 in 2021. The community (groups of counties) results have been drilled-down from the individual layers and the ones displaying a downward trend have been visualized in Figure \ref{fig:vaccine}. This illustration clearly shows how the \textbf{vaccination drive has become one of the reasons that has led to controlling the spread of COVID across US in the past few months}. This fact is also verified from independent sources that say how the administration of the vaccine has led to a \textit{decline in severe cases, hospitalizations and deaths} not only in the US, but some other parts of the world as well \cite{vaccine-1,vaccine-2}.

More layers and decoupling-based compositions addressing the analysis objectives listed below are being developed with a revised version of visualization that provides more interaction and choices.

\begin{itemize}
    \item What is the effect of traffic movement on new cases across major (or centrally connected) counties? How to choose a county for lockdown so that it has maximum impact? 
    \item Compare the increase/decrease in the number of new cases with respect to average education, per-capita income, mask usage and population density.
\end{itemize}

\section{Conclusions}
\label{sec:conclusions}

The success of any data analysis life cycle is predicated on our ability to: i) appropriately model the data set and automate schema generation as much as possible, ii) provide analysis alternatives using the model used, iii) map user-specified objectives to analysis expressions, iii) develop algorithms for computing expressions, preferably efficiently, and iv) ability to drill-down and visualize results for ease of understanding and interpretation.
This, as we know, is an iterative process.

In this paper, our focus has been to address the complete life cycle for aggregate analysis using the Multilayer Networks (or MLNs) as the underlying data model. In this paper, we demonstrate how to create EER diagrams for ``multi-entity, multi-feature" data sets. We use an algorithm that maps the EER diagram into MLNs of different kinds as appropriate. We have shown how user-specified objectives, in English, can be translated using heuristics based on keywords to aggregate analysis expressions for all types of MLNs. We have also demonstrated the applicability of the \textit{decoupling approach} for efficient analysis of complex data sets using multilayer networks. Drill-down and visualization of the results is an important but sometimes ignored component. We have shown how analysis results can be visualized using a general-purpose approach with a Covid-19 dashboard. 

We are working on further automating the translation of objectives into analysis expressions by using natural language processing and model characteristics. We are also working on developing decoupling-based efficient algorithms for aggregate analysis for centrality, substructure discovery, motif detection, to name a few, on MLNs. We are also broadening our analysis to include non-Boolean compositions for homogeneous MLNs, weighted graphs, and labeled graphs. This will further extend the expressive power of the MLN data model and the automation of the data discovery life cycle.

\begin{acknowledgements}
%If you'd like to thank anyone, place your comments here
%and remove the percent signs.
For this work, Drs. S. Chakravarthy and A. Santra were partially supported by NSF Grant 1955798 and Dr. Bhowmick was partially supported by NSF grant 1916084.
\end{acknowledgements}

% Authors must disclose all relationships or interests that 
% could have direct or potential influence or impart bias on 
% the work: 
%
% \section*{Conflict of interest}
%
% The authors declare that they have no conflict of interest.

% BibTeX users please use one of
%\bibliographystyle{spbasic}      % basic style, author-year citations
\bibliographystyle{spmpsci}      % mathematics and physical sciences
%\bibliographystyle{spphys}       % APS-like style for physics
%\bibliography{}   % name your BibTeX data base
\bibliography{./bibliography/somu_research,./bibliography/itlabPublications,./bibliography/itlabTheses,./bibliography/santraResearch,./bibliography/itlabPublications.original}

\end{document}